\documentclass[10pt, aps, prd, amssymb, amsmath, amsfonts, showpacs, floatfix, twocolumn, twoside, a4paper, superscriptaddress, longbibliography, nofootinbib]{revtex4-2}
\pdfoutput=1
\usepackage{newunicodechar}
\DeclareUnicodeCharacter{2009}{\,}

\usepackage{physics}
\usepackage{amsmath}	
\usepackage{amsthm}		
\usepackage{amssymb}	
\usepackage{natbib}
\usepackage{mathtools}
\usepackage{datetime}
\usepackage{graphicx}
\usepackage{color}
\usepackage{verbatim}
\usepackage{bm}
\usepackage{braket}
\usepackage{cancel}
\usepackage{xcolor}
\usepackage{slashed}
\usepackage{braket}
\usepackage{pifont}
\usepackage{amsfonts} 
\usepackage{float}
\usepackage{dsfont}
\usepackage[colorlinks=true, citecolor=midblue, linkcolor=midblue, urlcolor=midblue]{hyperref}
\usepackage[capitalise]{cleveref}
\usepackage{microtype}
\smallskipamount
\usepackage{feynmp}
\usepackage{letltxmacro} 
\usepackage{microtype}

\usepackage{setspace}

\usepackage{tikz,xcolor}
\definecolor{lime}{HTML}{A6CE39}
\DeclareRobustCommand{\orcidicon}{
	\begin{tikzpicture}
	\draw[lime, fill=lime] (0,0) 
	circle [radius=0.16] 
	node[white] {{\fontfamily{qag}\selectfont \tiny ID}};
	\draw[white, fill=white] (-0.0625,0.095) 
	circle [radius=0.007];
	\end{tikzpicture}
	\hspace{-2mm}
}
\foreach \x in {A, ..., Z}{
	\expandafter\xdef\csname orcid\x\endcsname{\noexpand\href{https://orcid.org/\csname orcidauthor\x\endcsname}{\noexpand\orcidicon}}
}

\settimeformat{ampmtime}

\definecolor{grey}{rgb}{0.4,0.4,0.4}
\definecolor{dullmagenta}{rgb}{0.4,0,0.4}
\definecolor{darkblue}{rgb}{0,0,0.4}
\definecolor{midblue}{rgb}{0,0,0.5}
\definecolor{midred}{rgb}{0.5,0,0}
\definecolor{orange}{rgb}{1,0.5,0}
\definecolor{lightbrown}{rgb}{0.75,0.5,0.25}
\definecolor{tan}{cmyk}{0.14,0.42,0.56,0}
\definecolor{djunglegreen}{cmyk}{0.99,0,0.52,0}
\definecolor{lightgreen}{rgb}{0,1,0}
\definecolor{olivegreen}{cmyk}{0.64,0,0.95,0.40}
\definecolor{midgreen}{rgb}{0.0,0.675,0.0}
\definecolor{darkgreen}{rgb}{0,0.5,0}

\settimeformat{ampmtime}

\usepackage{float}

\makeatletter
\let\oldr@@t\r@@t
\def\r@@t#1#2{%
\setbox0=\hbox{$\oldr@@t#1{#2\,}$}\dimen0=\ht0
\advance\dimen0-0.2\ht0
\setbox2=\hbox{\vrule height\ht0 depth -\dimen0}%
{\box0\lower0.4pt\box2}}
\LetLtxMacro{\oldsqrt}{\sqrt}
\renewcommand*{\sqrt}[2][\ ]{\oldsqrt[#1]{#2}}
\makeatother
\usepackage[normalem]{ulem}

\usepackage{soul}

\newcommand{\FirstAffiliation}{\affiliation{
Scuola Internazionale Superiore di Studi Avanzati (SISSA), via Bonomea 265, 34136 Trieste, Italy}
}
 
\newcommand{\SecondAffiliation}{\affiliation{Gravitation Astroparticle Physics Amsterdam (GRAPPA), 
    University of Amsterdam, 
    Science Park 904, 1098 XH Amsterdam, 
    The Netherlands
	}}

\newcommand{\AffiliationIFCA}{
\affiliation{Instituto de Fisica de Cantabria (CSIC-UC), Avda. Los Castros s/n, 39005 Santander, Spain}
}

\newcommand{\AffiliationINFNTS}{
\affiliation{Istituto Nazionale di Fisica Nucleare (INFN), Sezione di Trieste, via Valerio 2, 34127 Trieste, Italy}
}

\newcommand{\AffiliationIFPU}{
\affiliation{Institute for Fundamental Physics of the Universe (IFPU), via Beirut 2, 34151 Trieste, Italy}
}

\newcommand{\AffiliationINFN}{
\affiliation{INFN Sezione di Pisa, Polo Fibonacci, Largo B. Pontecorvo 3, 56127 Pisa, Italy}
}
 
\begin{document}

%
\title{Dark matter mounds from the collapse of supermassive stars:\\a general-relativistic analysis}
%
\author{Roberto Caiozzo}
\email{rcaiozzo@sissa.it}
\FirstAffiliation
\AffiliationINFNTS
\author{Gianfranco Bertone}
\SecondAffiliation
\author{Piero Ullio}
\FirstAffiliation
\AffiliationINFNTS
\AffiliationIFPU
\author{Rodrigo Vicente}
\SecondAffiliation
\author{Bradley J. Kavanagh}
\AffiliationIFCA
\author{Daniele Gaggero}
\AffiliationINFN
%

\begin{abstract}
\noindent
Recent work has highlighted the importance of a fully relativistic treatment of the dephasing of gravitational waves induced by dark-matter overdensities in extreme mass-ratio inspirals (EMRIs). However, a general-relativistic description of the dark matter phase-space distribution is currently available only for the case of a dark matter ``spike'' arising from adiabatic black hole growth. Here we develop a fully general-relativistic formalism for the more realistic scenario in which a supermassive stellar progenitor collapses to a black hole and produces a shallower dark matter overdensity, or ``mound''. We follow self-consistently the evolution of the supermassive star, its collapse, and the subsequent growth of the resulting black hole, together with the collisionless dark matter orbits. We find that in the regime where the collapse becomes non-adiabatic, the dark matter distribution function is significantly reshaped, with a clear depletion in the low-binding-energy region of phase space. Our results provide a more realistic prediction for the dark matter phase-space distribution around supermassive black holes, which is an essential step in using future EMRI observations to extract information about their formation history and the nature of dark matter.
\end{abstract}

\maketitle

\section{Introduction}
\label{sec:Introduction}
Dark matter (DM) is known to make up over 80\% of the matter content of our universe, yet its fundamental nature remains elusive~\cite{Bertone:2004pz,Bertone:2010zza,Bertone:2016nfn}. Our only hints of its properties come from its gravitational effects and from the absence of evidence in direct and indirect searches carried out so far~\cite{Arcadi:2017kky,Buckley:2017ijx,Bertone:2018krk,Cirelli:2024ssz}. On galactic scales, DM is expected to form extended halos characterized by a high central density~\cite{Dubinski:1991bm,Navarro_1996,Straight:2025udg}, and a structure that is sensitive to the assembly history of the host~\cite{Ludlow:2013bd,Schaller:2014uwa,Correa_2015,Calore:2015oya,Schaller:2015mua}. In particular, the formation and growth of supermassive black holes (SMBHs) at the centers of galaxies can substantially reshape the surrounding DM distribution, potentially generating large overdensities~\cite{Gondolo_1999,Ullio_2001,BERTONE2024116487}.

The precise structure of these central overdensities is of direct relevance for gravitational-wave (GW) astronomy. Extreme mass-ratio inspirals (EMRIs), in which a compact object spirals into a SMBH, are among the prime targets for future space-based GW detectors~\cite{Barack:2018yly,Berry:2019wgg,LISA:2022yao}. A dense DM environment around the SMBH modifies the orbital evolution through dynamical friction and accretion, leading to a cumulative dephasing of the GW signal \cite{Eda2013,Eda2015,Macedo2013,Barausse2014,Yue2019,Kavanagh2020,Becker2022,Speeney2022,Cardoso2022,Nichols2023,Karydas_2025a,Kavanagh_2025,Cardoso2020,Coogan2022,Cole2023,Cole2023Nature,Hannuksela2020}. Recent studies have highlighted the importance of a fully relativistic modeling of these effects, which requires a consistent treatment of both the EMRI dynamics and the DM phase-space distribution in curved spacetime~\cite{Vicente:2025gsg,Karydas:2025bkj}. 

The impact of SMBH growth on the central DM distribution has been most thoroughly studied in the idealized case where the black hole (BH) grows adiabatically from an initially infinitesimal seed at the center of a pre-existing halo, under the assumption that DM is cold and collisionless. In this limit, the change in the gravitational potential is slow compared to the orbital timescale of the DM particles, so that adiabatic invariants are conserved and the halo contracts, developing a steep inner overdensity, or ``spike''~\cite{Gondolo_1999}. The structure and robustness of such spikes have been analyzed in a variety of settings, including the effects of mergers, baryons, and self-annihilating DM~\cite{Ullio_2001,BERTONE2024116487}. This picture has been extended to the fully general-relativistic description of the phase-space distribution of collisionless DM around a Schwarzschild or Kerr BH formed by adiabatic growth in Refs.~\cite{Sadeghian_2013,Ferrer:2017xwm}. 

The formation and growth of SMBHs themselves are still poorly understood~\cite{Volonteri:2021sfo}. Recent observations of luminous quasars and galaxies at extreme redshifts, including with the James Webb Space Telescope, place standard ``light-seed'' channels based on stellar-remnant BHs and purely Eddington-limited growth under increasing tension~\cite{2023ApJ...955L..24G}. These findings motivate alternative formation pathways in which massive seeds with $M \sim 10^{4-6}\,M_\odot$ form at early times. One promising class of models involves the formation of supermassive stars (SMSs) in metal-poor, atomically cooling halos, which grow rapidly by accretion until they become unstable to general-relativistic collapse~\cite{Latif:2014cda,Chon:2018SMS}. In such ``direct-collapse'' scenarios, the gravitational potential in the central region can change on a timescale comparable to, or shorter than, the dynamical timescale of the innermost DM orbits. The adiabatic approximation then breaks down, and the final DM configuration can, in principle, be less steep~\cite{Ullio_2001}.

Recently, Ref.~\cite{Bertone_2025} adopted a semi-relativistic Monte Carlo treatment to investigate the DM distribution in a direct-collapse scenario where the SMBH is seeded by the collapse of a SMS at the center of an existing DM halo. In that framework, the formation of the overdensity proceeds in two stages. First, the SMS grows sufficiently slowly that the evolution of the central potential is adiabatic with respect to the DM orbits; then, the star undergoes a rapid collapse to a BH. This non-adiabatic phase erases part of the central spike, leading to a shallower DM overdensity, referred to as a ``mound''~\cite{Bertone_2025}. 

In this paper, we refine this calculation by developing a framework that is fully consistent with general relativity and that follows the evolution of the relativistic DM distribution function into its final, stable configuration. Our treatment explicitly includes the collapse phase of the SMS into a Schwarzschild BH, modeled using the Oppenheimer–Snyder (OS) solution of Einstein’s equations, and provides, for the first time, a self-consistent relativistic mapping between the pre- and post-collapse DM phase-space distributions.

The paper is structured as follows. In Sec.~\ref{sec:reldis}, we discuss the basics of relativistic distribution functions. In Sec.~\ref{sec:SG}, we review how to treat the evolution of the relativistic distribution function during the SMS growth. In Sec.~\ref{sec:OSR}, we show how to derive the final distribution function post-direct-collapse and how to correctly evolve the DM geodesics in the OS metric. The results of this procedure are shown in Sec.~\ref{sec:Results}. Finally we also show how the subsequent BH regrowth will eventually recover the adiabatic limit in Sec.~\ref{sec:Regrowth}.

In Appendix~\ref{sec:ICN} we show that our framework correctly reproduces previous results in the Newtonian limit, while in Appendix~\ref{sec:ICR} we analyze the limit of sudden metric changes in the comoving frame of the DM's particle ensamble and compare with earlier work.

We use units of $G=c=1$, and the mostly positive metric. Upper dots denote coordinate (Schwarzschild) time-derivatives. 

\section{Relativistic distribution function}
\label{sec:reldis}
We aim to derive the final DM distribution function and rest-mass density profile around a Schwarzschild BH, in a region where the the total DM mass is negligible compared to the black-hole mass. This allows us to neglect the back-reaction of the DM density evolution on the spacetime metric. This approximation is valid for the inner region of the DM spike considered here and was first developed for the case of adiabatic growth in \cite{Sadeghian_2013}. We start by reviewing the basics of relativistic distribution functions and presenting some key expressions. Taking $\dd N$ to be the number of collisionless particles crossing the spacelike 3-volume element $\dd V$ at the event $x$, with 4-velocity $u^\alpha$ in the range $\dd V_u$, the invariant distribution function $f(x,u^i)$ is defined through~\cite{Lindquist:1966igj,RST}
\begin{equation}
    \dd N = f(x,u^i)(-u^\alpha s_\alpha)\,\dd V \dd V_u\,,
\end{equation}
with $s_\alpha$ the future-pointing unit normal to $\dd V$, and $\dd V_u$ a 3-surface element satisfying the constraint $u^\alpha u_\alpha=-1$.

From Jeans' theorem~\cite{Jeans:1915, Fackerell, Rioseco_2017}, the distribution function of a steady-state, collisionless system depends on the phase-space coordinates only through the integrals of motion. For spherically symmetric systems we can write
\begin{equation}
    f(x,u^i)\equiv\, f(\mathcal{E},\mathcal{L})\,,
\end{equation}
with $\mathcal{E}$ and $\mathcal{L}$ the specific energy and angular momentum of the particles. 
The main targets of this work are the DM distribution function and the associated particle number density profiles. We will be interested in computing density profiles over spacetime regions that are static, and so admit a timelike Killing vector field, $t^\alpha\equiv\delta_t^\alpha$.
The number density profiles on $\Sigma\equiv\{t=\text{const}\}$ hypersurfaces are obtained through
\begin{equation}
\begin{aligned}
    n_\Sigma\equiv -\int f\, u^\alpha s_\alpha\,\dd V_u&  \\=\frac{4\pi}{r^2\sqrt{g_{rr}}}\int \mathcal{L}\, f(\mathcal{E}, \mathcal{L})\, \frac{u^t}{|u^r|}\, \dd \mathcal{E}\, \dd \mathcal{L} \,, \label{eq:n_density}
\end{aligned}
\end{equation}
with $r$ the areal radius. This can then easily be converted to the rest mass density profiles $\rho_0=\mu \,n_\Sigma$ as computed in~\cite{Sadeghian_2013,Ferrer:2017xwm} by multiplying by the DM rest mass $\mu$. To obtain the right-hand-side we used~$s^\alpha=t^\alpha/\sqrt{-g_{tt}}$, the momentum element $\dd V_u= (\sqrt{-g}/\mathcal{E})\, \dd u^r \dd u^\theta \dd u^\varphi$  (in coordinate basis), the Jacobian
\begin{equation}
    J\equiv \Big|\det\, \frac{\partial(\mathcal{E},\mathcal{L},\mathcal{L}_z)}{\partial(u^r,u^\theta,u^\varphi)}\Big|=\frac{(-g_{tt})\,r^6\sin^2\theta\,|u_r| \, |u^\theta|}{\mathcal{L}\, \mathcal{E}}\,,
\end{equation}
and we integrated over the third component of the angular momentum $\mathcal{L}_z$ using $\int \dd\mathcal{L}_z/|u^\theta|=\pi\,r^2 \sin \theta$, from $|u^\theta|=\sqrt{\mathcal{L}^2-(\mathcal{L}_z/\sin\theta)^2}/r^2$.\footnote{We note that the only dependence of the integrand in $\mathcal{L}_z$ arises through $J^{-1}$.} We included a factor of~4 in the change of variables to account for the signs of $u^\varphi$ and $u^r$. 

By integrating $\dd N$ over the spatial volume on $\Sigma$ with $(\mathcal{E},\mathcal{L})$ held fixed, using the induced volume element $\dd V=\sqrt{g/g_{tt}}\,\dd r\, \dd\theta\, \dd\varphi$, it is straightforward to show
\begin{equation}
     f(\mathcal{E},\mathcal{L}) = \frac{1}{8 \pi^2\mathcal{L} \, T_r(\mathcal{E},\mathcal{L})}\frac{\dd N}{\dd \mathcal{E} \dd \mathcal{L}}\,,
     \label{fEL}
\end{equation}
with $T_r\equiv 2\int (u^t/u^r) \dd r$ the radial period as measured by stationary observers at infinity. 

In general, a DM distribution will exhibit some time dependent fine-structure in response to a sudden change in the spacetime metric. Nevertheless, if spherical symmetry is preserved throughout the collapse, one can write
\begin{equation}
    \frac{\dd N}{\dd \mathcal{E} \dd \mathcal{L}}=8\pi^2 \mathcal{L}\oint_\Sigma f(t,r,\mathcal{E},\mathcal{L},\nu)\,\frac{u^t}{|u^r|}\,\dd r\,,
    \label{dnde}
\end{equation}
with $\nu \equiv \mathrm{sign}\,u^{r}$, and where the contour integral runs over the entire orbit (i.e., including both $\nu=\pm 1$ branches). Note that Eq.~\eqref{dnde} does not rely on any assumption of the phase mixing discussed in Sec.~\ref{sec:OSR}. The expression simply counts the number of particles crossing the hypersurface per unit $\dd\mathcal{E}$ and $\dd\mathcal{L}$, and therefore remains valid even when the distribution function exhibits explicit time dependence, provided spherical symmetry is preserved. Crucially, for a collisionless system evolving in a static metric, the energy and angular momentum of each particle are conserved. As a result, $\dd N/(\dd\mathcal{E},\dd\mathcal{L})$ is independent of the choice of the $t=\mathrm{const}$ hypersurface, as long as the metric is unchanged over the region explored by the corresponding orbits. Hence, for any such choice of $t$, Eq.~\eqref{dnde} yields the same result.

\section{Supermassive Star Growth}
\label{sec:SG}
We model the evolution of the DM distribution function during the SMS adiabatic growth, assuming as reference model that the initial DM distribution is well described by an NFW profile~\cite{Navarro_1996} with a virial mass of $M_\text{vir} = 10^7\,M_\odot$ at redshift $z_{\rm f} = 15$. For these fiducial values, the corresponding halo concentration is $c = 3$~\cite{Correa_2015}. 

Due to its very low compactness, the initial DM halo is well described by Newtonian mechanics. Assuming isotropy at formation, the initial DM distribution function is obtained through the Eddington inversion procedure~\cite{1987gady.book.....B}:
\begin{equation}
    f_\text{nfw}({E})=\frac{1}{\sqrt{8} \pi^2} \frac{\mathrm{d}}{\mathrm{d} {E}} \int_0^{{E}} \frac{\mathrm{d} \Phi_{\text{nfw}}}{\sqrt{{E}-\Phi_{\text{nfw}}}} \frac{\mathrm{d} n_{\text{nfw}}}{\mathrm{~d} \Phi_{\text{nfw}}}\,,
\end{equation}
where $n_{\text{nfw}}(r)$ and $\Phi_{\text{nfw}}(r)$ are the radial number density and potential for the NFW profile, $E=\Phi-v^2/2$ is the non-relativistic specific energy. The distribution function is set to zero for unbound particles with $E\leq0$.
Note that in the non-relativistic limit $\mathcal{E}\approx 1-E$.

An SMS could form at the center of the halo through a direct-collapse pathway in which H$_2$ cooling is suppressed, allowing the gas to collapse monolithically and feed the protostar at an accretion rate of $\sim0.1\,M_\odot\,\text{yr}^{-1}$~\cite{Hosokawa_2013}. The star will continue to grow until it reaches $\approx 10^5\,M_\odot$, at which point general-relativistic instability triggers direct collapse to a BH~\cite{Hosokawa_2013,Chon_2018,Umeda_2016}. The growth timescale of the star, $t_{\rm growth}=10^6\,$yr, is much longer than both its characteristic hydrodynamical timescale and the dynamical timescale of the DM halo at the star's radius $t_{\rm dyn}=\sqrt{(3\pi)/(32\bar{\rho})}\sim10^3\,$yr, where $\bar{\rho}$ is the average density of the NFW for radii smaller than the star's radius. In the following we can then safely assume an adiabatic evolution for the system. 

We approximate the interior of the pre-collapse star as having uniform density. This is a reasonable approximation, particularly for radii $r \lesssim 0.1,R_\star$, as shown in Fig.~3 of Ref.~\cite{Hosokawa_2013}, where $R_\star$ denotes the stellar radius. This choice also ensures consistency with the Oppenheimer–Snyder model used to describe the direct collapse. Motivated by the results of Ref.~\cite{Hosokawa_2013}, we fix the stellar density to $\rho_s = 1\,\mathrm{g\,cm^{-3}}$ and determine $R_\star $ such that the star’s mass at collapse is $M = 10^5\,M_\odot$. The SMS is then modeled as an isotropic perfect fluid star whose metric solves the Tolman–Oppenheimer–Volkoff equation~\cite{Oppenheimer:1939ne}):
\begin{equation}
    \dd s^2 = -A\, \dd t^2 +\frac{\dd r^2}{1-\frac{2m(r)}{r}}+r^2\dd\Omega^2\,,
\end{equation}
with the enclosed mass $m(r)=M[\min(r,R_\star )/R_\star ]^3$, and
\begin{equation}
    A=\begin{cases}
        \frac{1}{4}\left[3\sqrt{1-\dfrac{2 M}{R_\star }}
-\sqrt{1-\dfrac{2 m(r)}{r}}\;\right]^2, \quad r\leq R_\star \\ 
        1-\dfrac{2M}{r}\,, \quad r>R_\star \,.
    \end{cases}
\end{equation}

Since the adiabatic invariants are conserved, we compute the distribution function of the DM after the SMS formation using the method developed in Ref. \cite{1972GReGr...3...63P,1995ApJ...440..554Q} and adopted in Refs.~\cite{Gondolo_1999,Bertone_2025}, or more precisely its general-relativistic extension~\cite{Sadeghian_2013,Ferrer:2017xwm}.

For spherically symmetric static spacetimes (describing both the initial DM halo and the final pre-collapse SMS), the relativistic adiabatic invariants~\cite{witzany2022actionanglecoordinatesblackholegeodesics,Landau_adiabatic} force the conservation of angular momentum and radial action. These can be written as
\begin{align}
     \mathcal{L}&= r^2\frac{\mathrm{d}\varphi}{\mathrm{d}\tau}\,,\\  \mathcal{I}_{r}&=\oint u_r \, \dd r=2\int \sqrt{g_{rr}\left(\frac{\mathcal{E}^2}{g_{tt}}-1-\frac{\mathcal{L}^2}{r^2}\right)}\;\mathrm{d}r\,,
\end{align}
where $\tau$ is the DM particle's proper time.
In the Newtonian limit, they reduce to
\begin{align}
    \mathcal{L}&\approx L\equiv r^2\frac{\mathrm{d}\varphi}{\mathrm{d}t}\,,\\
    \mathcal{I}_r&\approx I_{r,{\rm N}}\equiv 2\int\sqrt{2[\Phi(r)-E]-\frac{L^2}{r^2}} \;\mathrm{d}r\,.
\end{align}

The distribution function after the SMS has formed can be retrieved using $\mathcal{I}_r(\mathcal{E}_{s},\mathcal{L})\approx I_r(E_\text{nfw},L)$ and $\mathcal{L}=L$ to map the initial and final distribution function by writing
\begin{equation}
    {f}_{s}(\mathcal{E}_{s},\mathcal{L}) \approx f_\text{nfw}(E_\text{nfw}(\mathcal{E}_s,\mathcal{L})),
\end{equation} where the subscript $s$ denotes quantities referring to the fully formed SMS.


\section{Oppenheimer-Snyder Collapse}
\label{sec:OSR}
The Oppenheimer-Snyder (OS) solution~\cite{OS} describes the gravitational collapse of a pressureless sphere of dust with uniform rest mass density. Its metric is 

\begin{equation}
    \dd s^2 = -\dd\eta^2 + B^2(\chi) \bigg(\frac{\dd R^2}{1 - \frac{2 m(R)}{R}} + R^2 \dd \Omega^2 \bigg)\,, 
\end{equation}
where
\begin{align}
    \eta(\chi) &= \frac{1}{2} \sqrt{\frac{R_\star ^3}{2M}} \left( \chi + \sin \chi \right)\,,\\
    B(\chi) &= \frac{1}{2} \left( 1 + \cos \chi \right) = r/R\,.
\end{align}

The spacelike hypersurface $\eta=0$ defines the beginning of the collapse. Within the star ($R< R_\star $) the coordinate $\eta$ gives the proper time of comoving observers with the collapsing (baryonic) matter, while at the surface ($R=R_\star $) it gives the proper time of radially free-falling observers in Schwarzschild starting from rest at $r=R_\star $ and $t=0$. The coordinate $R$ labels the worldline of each of those observers; it is chosen such that at $\eta=0$ it coincides with the areal radius $r$. For later convenience, it will be useful to note that 
\begin{equation} 
\frac{\dd B}{\dd\eta}=\frac{\dd B}{\dd \chi}\frac{\dd \chi}{\dd \eta}=-\sqrt{\frac{2M}{R_\star ^3}(B^{-1}-1)}\;. 
\end{equation}
\subsection{Matching SMS to OS collapse}
We assume that the collapse is initiated at a spacelike hypersurface $\Sigma_0$, defined by $t=0^-$ in the SMS spacetime and $\eta=0^+$ in the OS collapse spacetime. For the resulting spacetime to be a solution of Einstein equations, the matching is required to satisfy Israel's junction conditions~\cite{Israel1,chu2021}. These are (i) the continuity of the induced metric across $\Sigma_0$, i.e., $h_+=h_-$, and (ii) the jump in the extrinsic curvature
\begin{equation}
    [K_{ij}]=8\pi \left(S_{ij}-\frac{h_{ij}}{2}S\right)\,,
\end{equation}
where $S_{ij}$ is the so-called surface stress tensor, which is the pull-back of the stress-energy tensor integrated over a small region around
the hypersurface $\Sigma_0$. The condition (i) is trivially satisfied in our case. It is also not hard to show that $K_{ij}=0$ for both $t=0^-$ and $\eta=0^+$; the condition (ii) implies then $S_{ij}=0$ (no matter shell at $\Sigma_0$). The unphysical discontinuity of $T_{\alpha \beta}$ across the hypersurface (with the isotropic pressure vanishing suddenly) does not pose a problem to the Einstein equations sector. 

Since there is no discontinuity of any first derivative of the metric across $\Sigma_0$, the geodesics are smooth (including along the normal direction to $\Sigma_0$). From $(r,\theta,\varphi)=(R,\theta,\varphi)$ at $\Sigma_0$, we have $(u_r, u_\theta, u_\varphi)=(u_R, u_\theta, u_\varphi)$ there. The identity $u^\alpha u_\alpha=-1$ implies then $(u_t)^2=A(R)(u_\eta)^2$ at the hypersurface.

\subsection{Connecting the pre- to post-collapse\\DM distribution functions}
The distribution function of collisionless systems is conserved along geodesics, according to the relativistic Liouville theorem~\cite{Liouville1838,Lindquist:1966igj}:
\begin{equation}
    f(x,u^i)|_{\zeta}= {\rm constant}\,,
\end{equation}
with $\zeta(\tau)$ denoting the DM particle geodesics; notably, this holds even when energy is not conserved. We follow then the evolution of the distribution function up to $t_f$,
\begin{equation}
    f(x,u^i)|_{t=t_f}=f_s\big(\mathcal{E}_s(r_f,u^r_f,\mathcal{L}),\mathcal{L}\big)\;,
\end{equation}
with $t_f$ sufficiently in the far future such that the radius of the star as a function of the coordinate time $r_\star (t)$ has shrunk to $r_\star (t_f)\leq 4M$, so that all stable bound orbits remain in the static exterior region of the OS metric.\footnote{Note that there are no stable Schwarzschild orbits with periapsis smaller than $4M$} (For the numerical computations we will choose $t_f$ such that $r_\star(t_f)=4M$). The subscript $f$ denotes the values on the hypersurface $t=t_f$. Note here that DM is collisionless and that after $t=t_f$ the metric is static for all relevant orbits. In the following section we discuss how to find $\mathcal{E}_s(r_f,u^r_f,\mathcal{L})$ by solving the geodesics of OS from $t=t_f$ backwards to $t=0$. 

Crucially, in collisionless systems, for orbits confined to static regions of the metric, $\dd N/(\dd\mathcal{E}\,\dd\mathcal{L})$ remains constant in time even if the distribution function still exhibits time dependence. We can then compute Eq.~\eqref{dnde} at $t=t_f$ and extend this computation to all $t>t_f$. 

During and after the collapse, the DM particles are in a non-uniform orbital-phase configuration, as they inherit their initial orbital phase from the pre-collapse SMS. The distribution is therefore explicitly time dependent, and Jeans' theorem does not apply. As the DM particles evolve in the changing metric of the collapsing star the system experiences violent relaxation, undergoing an intricate orbital mixing process that generates intermingled filaments in phase space~\cite{1967MNRAS.136..101L,chavanis2002statisticalmechanicsviolentrelaxation}. Within a few dynamical times (comparable to the orbital periods of the relevant particles, $\sim \sqrt{R_s^3/GM}\sim 10 \;\text{hr}$) after the collapse has effectively ended, or $t>t_f$, phase-space mixing drives the system toward an approximate coarse-grained steady state equilibrium~\cite{Tremaine1986HFunctions,1987gady.book.....B,Merrall_2003} (see also~\cite{RiosecoSarbach2024,RiosecoSarbach2020_PhaseSpaceMixing} for the general-relativistic case). 

After sufficient time has passed, we can then use Eq.~\eqref{fEL} together with the value of Eq.~\eqref{dnde} computed at $t=t_f$, to write the (phase-mixed) coarse-grained post-collapse distribution function as

\begin{equation}
 \begin{aligned}  f_c(\mathcal{E}_c,\mathcal{L})=\frac{1}{T_r(\mathcal{E}_c,\mathcal{L})}\oint  f(x,u^i)|_{t=t_f}\,\frac{u_f^t}{u_f^r}\,\dd r_f\\=\frac{1}{T_r(\mathcal{E}_c,\mathcal{L})}\oint  f_s\big(\mathcal{E}_s(r_f,u^r_f,\mathcal{L}),\mathcal{L}\big)\,\frac{u_f^t}{u_f^r}\,\dd r_f\,,
 \label{coarsef}
 \end{aligned}
\end{equation}
where the subscript $c$ refers to the values long after the collapse. Note that, after the SMBH has formed, we set the distribution function of particles in the loss-cone (as computed in~\cite{Sadeghian_2013}) to zero. In practice, these particles will be captured by the SMBH, resulting in a small (negligible) increase in the mass of the central object.
\subsection{Geodesics of OS from $t=t_f$ backward to $t=0$}
The final step needed to obtain the distribution function after the OS collapse is to find $\mathcal{E}_s(r_{ f},u^r_f,\mathcal{L})$. 
Our choice $r_\star(t_f)=4M$ (see previous section) corresponds to
\begin{equation}
    t_f=\int_{4M}^{R_\star } \frac{U^t}{U^r} \,\dd r = \int_{4M}^{R_\star }\frac{\sqrt{1-\frac{2M}{R_\star }}}{(1-\frac{2M}{r})\sqrt{\frac{2M}{r}-\frac{2M}{R_\star }}}\,\dd r \,,
\end{equation}
with $U_\alpha$ the 4-velocity of a point at the SMS surface.

DM particles in stable bound orbits after collapse can be divided in three classes: (i) those that remain in the exterior of the star at all times, (ii) those that start in the exterior and cross the star surface twice, and (iii) those that start inside and cross the star surface once. Because the star surface is free falling, it is clear that geodesics crossing it outwards will \emph{not} cross it inwards again (see Appendix~\ref{A} for a more thorough discussion). 

The class (i) is trivial: the energy and angular momentum of those particles is conserved through collapse. In particular: $\mathcal{E}_s=\mathcal{E}_c$. 

For both classes (ii) and (iii) there is an instant $t_{\rm out}$ when the radial position of the DM particle is such that $r_{\zeta}(t_{\rm out})=r_\star(t_{\rm out})$ and $\dot{r}_\zeta(t_{\rm out})>\dot{r}_\star(t_{\rm out})$, i.e., when the particle exits the star. In the exterior of the star we evolve the geodesics in the Schwarzschild coordinates $(t,r)$, while in the interior we use $(\eta,R)$. The OS coordinates of the crossing event are 
$R_{\rm out}=R_\star$ and $\eta_{\rm out}=\eta(\chi_{\rm out} )$, with $\chi_{\rm out}=\cos^{-1}(2 (r_\text{out} /R_\star )-1)$. In the OS coordinate basis the DM particle 4-velocity is then

\begin{equation}
\begin{aligned}
    u^\eta&=\partial_\alpha \eta\, u^\alpha=-U_\alpha u^\alpha\,, \\
    u^R&=\frac{\sqrt{(u^{\eta})^2-1-(\mathcal{L}/r)^2}}{\sqrt{g_{RR}}}\impliedby u^\alpha u_\alpha = -1\,.
\label{uR}
\end{aligned}
\end{equation}

Inside the star the geodesic of the radius of the particle is evolved by using the geodesic equation 

\begin{equation}
    \begin{aligned}
 \frac{\dd u^R}{\dd \tau}&=\left(R-\frac{2MR^3}{R_\star^3}\right)\left(u^\varphi \right)^2\\
 &\qquad -\frac{2MR\;(u^R)^2}{R_\star ^3-{2MR^2}}-2\frac{\dd B}{\dd\eta}\frac{u^Ru^\eta}{B}\,,
\end{aligned}
\end{equation} 
while $\eta$ is evolved in parallel by using Eq.~\eqref{uR}. We solve these two equations numerically up to $\eta=t=0$ in case (iii), or up to when the DM particles reach $R=R_\star (\eta_{\rm in})$ again in case (ii). 

In case (iii) the energy before the collapse is finally found through
\begin{equation}
    \mathcal{E}_s=-u_t{\,}|_{t=0}=-\sqrt{A}\, u_\eta{\,}|_{\eta=0}\,,
\end{equation}
while in case (ii) is obtained by
\begin{equation}
    \mathcal{E}_s=-u_t{\,}|_{\eta=\eta_{\rm in}}\,,
\end{equation}
where $u_t$ can be obtained by using the four velocity transformation
\begin{equation}
    u^r=\partial_\alpha r\, u^\alpha =B\,u^R+R\frac{\dd B}{\dd \eta}\,u^\eta
\end{equation}
using $r_\mathrm{in}=B(\eta_\mathrm{in})\,R_\star $. Then using the normalization of 4-velocity we obtain
\begin{equation}
\mathcal{E}_s=\sqrt{\left(1+\frac{\left(u^r\right)^2}{1-\frac{2M}{r_\mathrm{in}}}+\frac{\mathcal{L}^2}{r_\mathrm{in}^2}\right)\left(1-\frac{2M}{r_\mathrm{in}}\right)}\,.
\end{equation}

\section{Results}
\label{sec:Results}
\begin{figure}[t]
\includegraphics[width=0.45\textwidth]{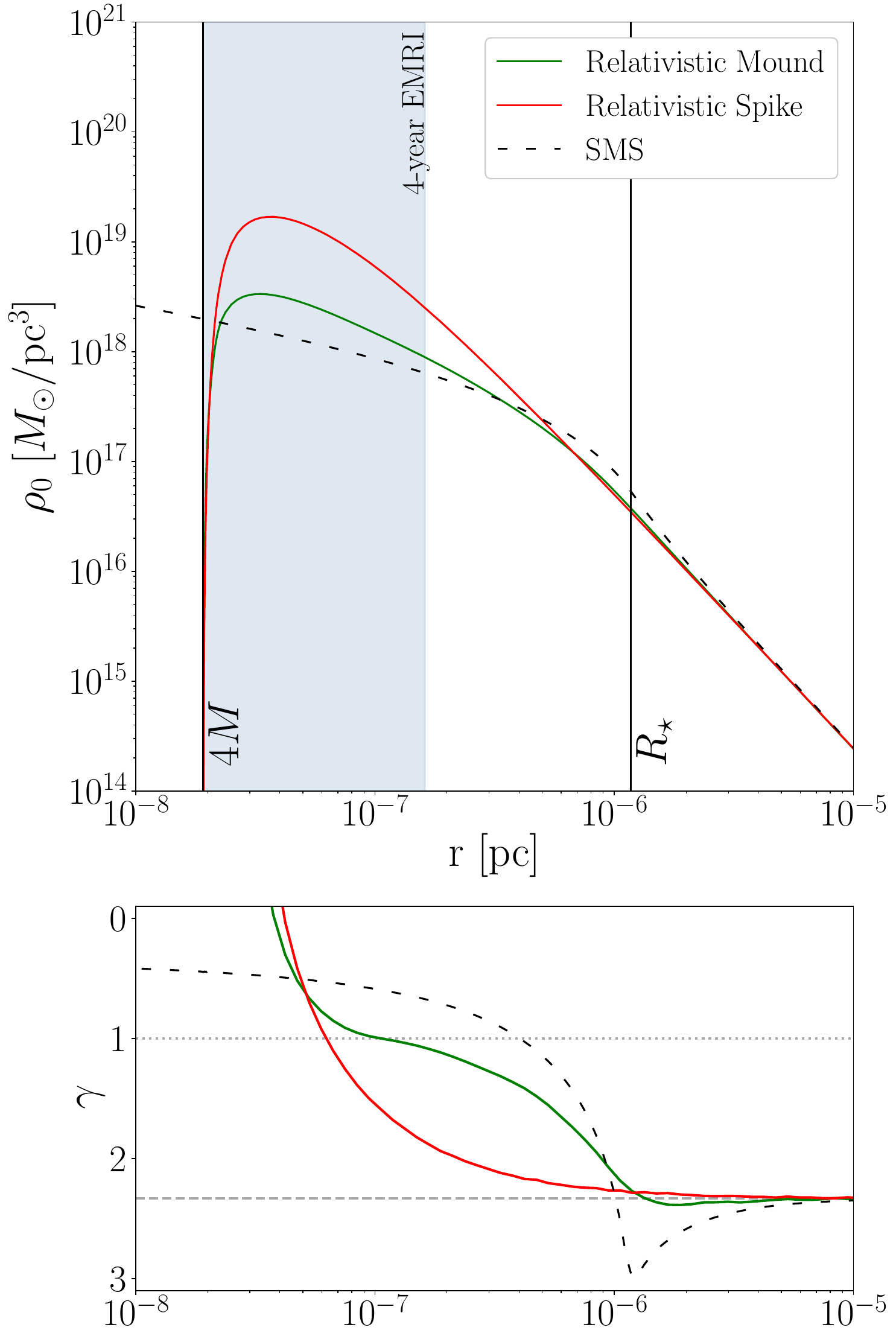}
\caption{\textbf{Top panel}: DM rest mass density profile
pre- and post- SMS collapse; these are labeled, respectively, by SMS and Relativistic Mound. The case of a SMBH growing adiabatically from a light seed to a mass of $10^5\,M_\odot$ is also shown for comparison and it is labeled Relativistic Spike. \textbf{Bottom panel}: Profile slope, $\gamma\equiv-\dd \log \rho/\dd r$, for each case.}
\label{comp}
\end{figure}
\begin{figure*}[t]
    \centering
    \includegraphics[width=0.48\textwidth]{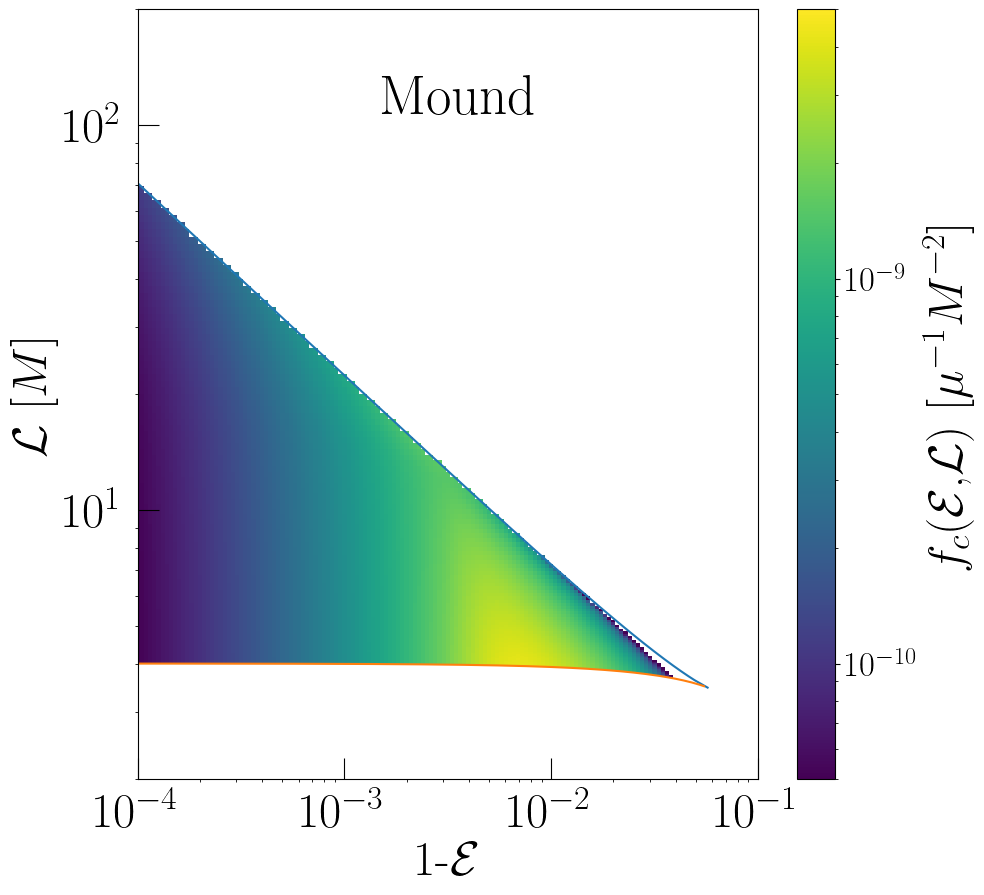}
    \includegraphics[width=0.48\textwidth]{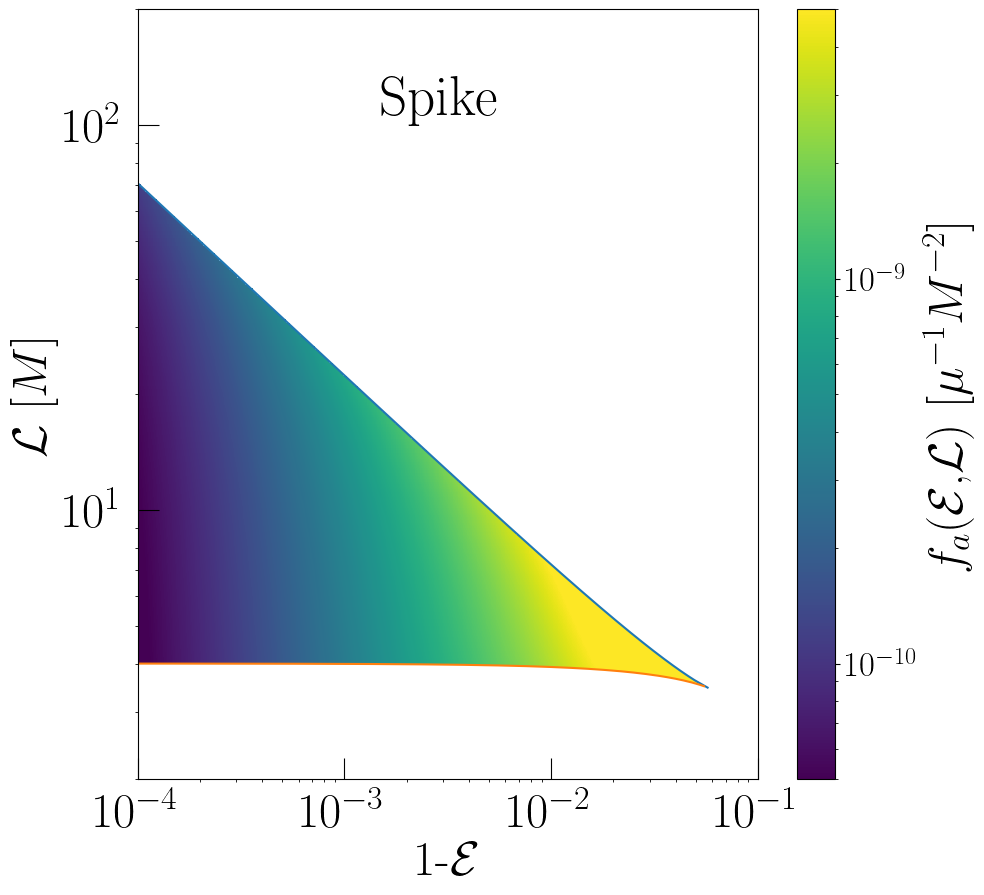}
    \hfill
    
    \caption{
    \textbf{Left}: Coarse-grained distribution function of DM after OS collapse projected in the phase space $(\mathcal{L},1-\mathcal{E})$. \textbf{Right}: Same quantity but for a spike grown adiabatically. In both panels, the top blue curve corresponds to circular orbits and the bottom orange curve corresponds to the minimum angular momentum required to avoid capture by the BH for the given energy.
    }
    \label{fig:DF_comparison}
\end{figure*}
\begin{figure*}[t!]
    \centering
    \includegraphics[width=0.48\textwidth]{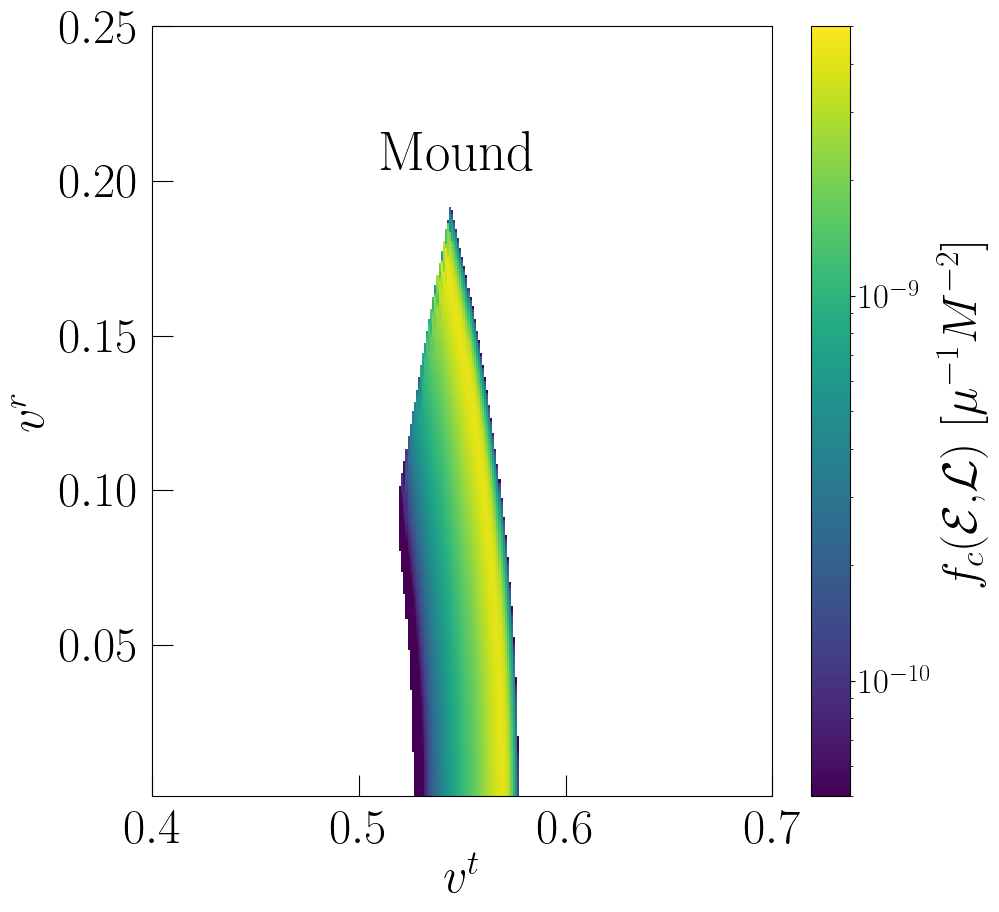}
    \includegraphics[width=0.48\textwidth]{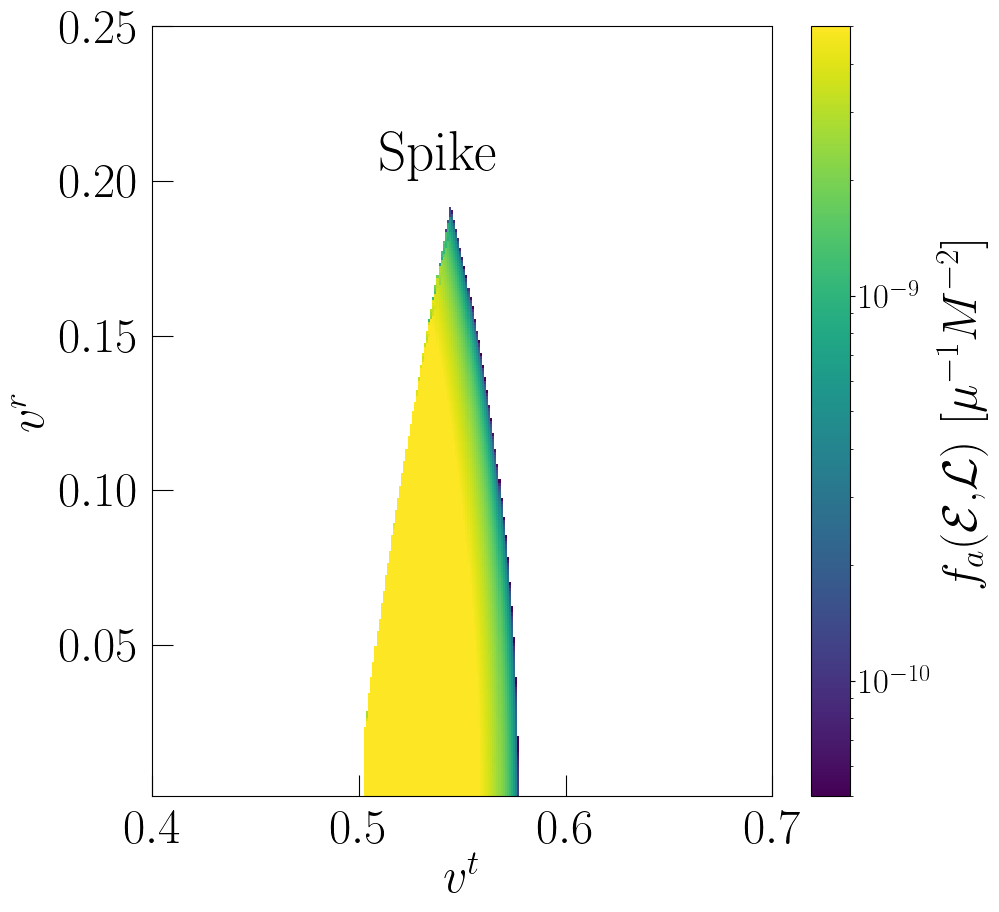}
    \hfill
    
    \caption{
        \textbf{Left}: Distribution function of DM particles in the projected phase space of velocities (radial and tangential) as measured in the LIF at $6M$ for a spike following OS collapse. 
        \textbf{Right}: Same quantity but for a spike grown adiabatically.
    }
    \label{fig:LIF_comparison}
\end{figure*}
In Fig.~\ref{comp} we show DM rest mass density profiles obtained as the integral of the phase space distribution function, see Eq.~\eqref{eq:n_density}, multiplied by the DM rest mass $\mu$ after the SMS formation and its collapse into a BH within the OS model. We refer to this DM density profile as a relativistic mound. The growth of the central object has sharply enhanced the DM density (the initial NFW density, not shown in the plot, is about $3\times10^9 \,M_\odot/\text{pc}^3$ at $4M$). The plot focuses on radii which are relevant for GW signals from EMRIs which are within the size $R_\star$ of the SMS before its collapse, the shaded region in the plot highlights the
range of radii probed by an extreme mass-ratio inspiral
with companion mass $10 \, M_\odot$, starting 4 years
before the merger. The collapse causes a redistribution of DM within radii which are a few times larger then $R_\star$, and all the way to the region relevant for EMRIs, with a sharp cutoff at twice the Schwarzschild radius. 

Such modification is indeed sensitive to the formation history of the SMBH. As an example, for comparison, we show the result in the case where the final configuration is reached assuming an adiabatic formation of a BH of equal mass, with the associated distribution function $f_a$, derived by applying the same formalism for adiabatic contraction as discussed in Sec.~\ref{sec:SG} (this is the same idealized case discussed for the first time in Ref.~\cite{Sadeghian_2013}). We refer to this DM rest mass density profile as a relativistic spike. The normalization of the inner density in this latter case is noticeably larger than for our result close to the BH, about a factor of 5 at the innermost stable circular orbit ($r=6M$). This is in agreement with general considerations that predict a smaller focusing of the DM density for processes that involve a rapid change in the underlying potential well. 

In the lower panel of Fig.~\ref{comp}, we stress differences in the logarithmic slope of the profile, which is also a relevant parameter when computing the dephasing of EMRIs due to accretion and dynamical friction. The difference in density profiles stems from the different redistribution in phase space of the DM particles in the two BH formation models. This is illustrated in Fig.~\ref{fig:DF_comparison} where we show a color coding of the magnitude of the distribution functions in the projected parameter space $(\mathcal{L},1-\mathcal{E})$. Here $1-\mathcal{E}$ measures how strongly bound the particles are to the central object and, together with $\mathcal{L}$, determines the peri- and apoapsis of the DM orbits. In our reference model (left panel), there is a depletion in the region associated to orbits with apoapsis close to the BH, namely those with low $\mathcal{E}$, more pronounced at larger values of $\mathcal{L}$, namely the circular orbit limit. In this limit, for a fixed energy $\mathcal{E}$, the apoapsis is minimized, which in turn tends to maximize the initial energy required to reach these final orbits. The depletion is due to the fact that this region of parameter space corresponds to pre-collapse energies associated with unbound orbits, $\mathcal{E}_s \geq 1$. This is in sharp contrast with the adiabatic case where small radii circular orbits persist, see Fig.~\ref{fig:DF_comparison}, right panel. These results are consistent with previous semi-relativistic calculations under the naïve assumption of an instantaneous collapse, however, the more realistic treatment of the dynamics of collapse mitigates the suppression with respect to the adiabatic case, see Appendix~\ref{sec:ICR}.

Such details in the underlying distribution function are relevant for future GWs observation because the different density and velocity distribution close to the BH will inevitably impact on accretion and dynamical friction within the 4 year EMRI inspiral~\cite{Vicente:2025gsg}. In particular Fig.~\ref{fig:LIF_comparison} illustrates the differences in the distribution function for the two cases in the projected parameter space of radial and tangential velocities of DM particles as perceived by an observer in the local inertial frame (LIF) associated to the rest frame of the DM particle ensemble at $6M$.

The impact on EMRI waveforms will be investigated in upcoming work along with a discussion on potential insights into the formation history of SMBHs.

\section{Regrowth}
\label{sec:Regrowth}
\begin{figure}
\includegraphics[width=0.45\textwidth]{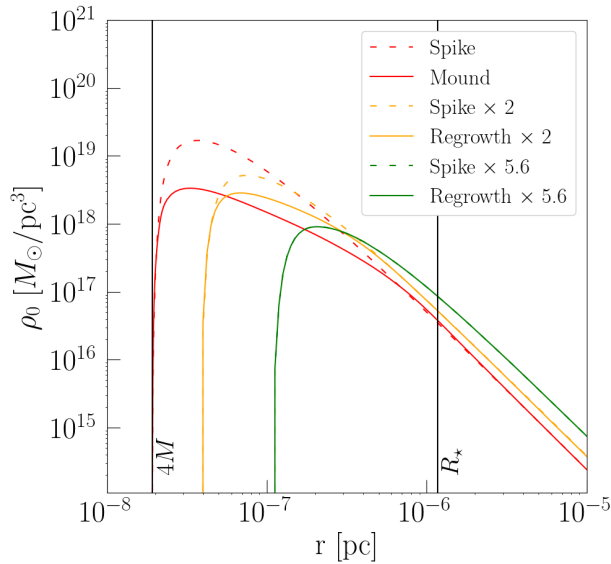}
    \caption{DM rest mass density profiles after the SMS collapse (Mound) and various factors of regrowth $n=M_f/M$ (Regrowth $\times$ $n$). The adiabatic equivalents for the initial BH mass (Spike) and each regrowth factor (Spike $\times$ $n$) are also shown as dotted lines.}
\label{reg}
\end{figure}

After the SMS collapse, the resulting BH might further increase its mass through accretion, leading to a second phase of adiabatic growth. Once again this can be accounted for through the relativistic adiabatic invariants and the associated evolution of the distribution function. In Fig.~\ref{reg} we show how the density profile responds to the subsequent regrowth of the central object, considering as sample cases final BH masses which are 2 and 5.6 times the initial mass respectively. We also show the equivalent spike grown adiabatically from 0 mass in each case for comparison. As expected this slowly erases some of the imprints of the phase of rapid collapse. Our scenario is still appreciably different from the purely adiabatic limit when the BH mass doubles, while for a regrowth of 5.6 the two cases coincide, as explained below.

\begin{figure}
\includegraphics[width=0.45\textwidth]{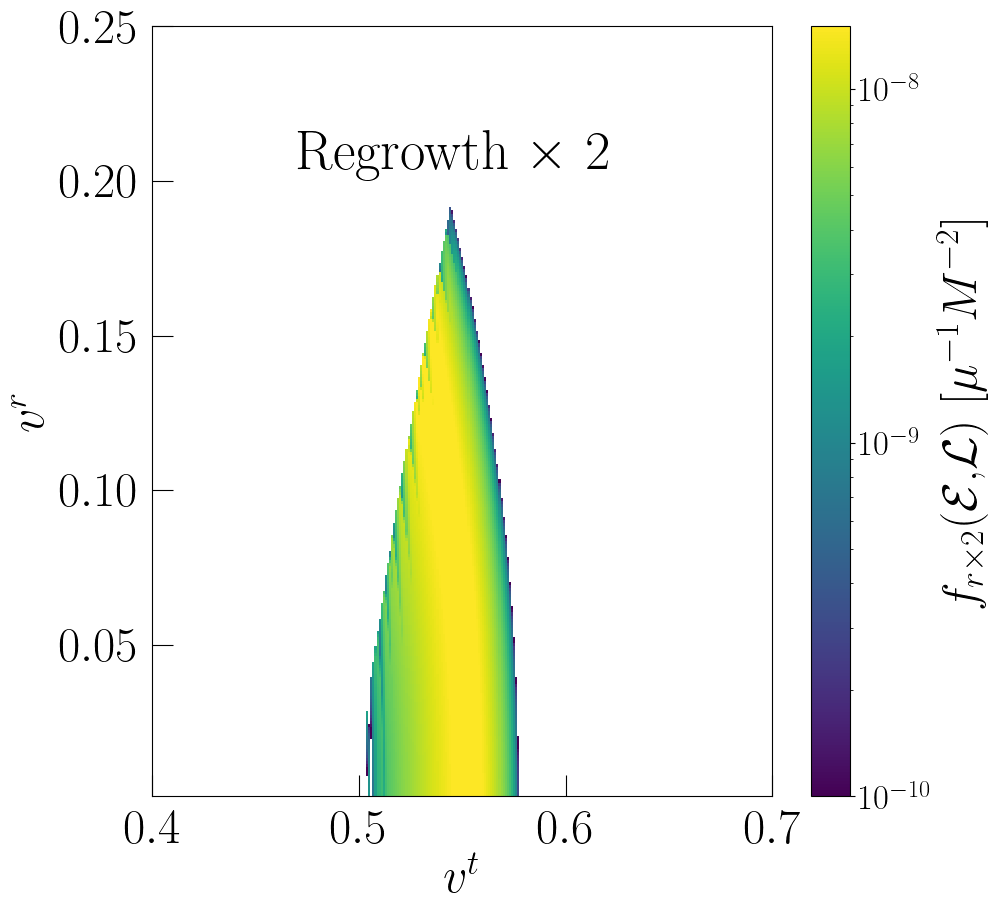}
    \caption{Distribution function of DM particles for their radial and tangential velocity, as perceived by an observer in the LIF at $6M$, for a spike post OS collapse and a subsequent regrowth by a factor 2.}
\label{r3rs}
\end{figure}
Fig.~\ref{r3rs} shows the same projection as in Fig.~\ref{fig:LIF_comparison} in the case where the BH mass grows to twice its initial size, with the associated distribution function $f_{r\times 2}$. The high density region of the distribution function at $3\,R_\star $ slowly drops towards the low tangential velocities associated to orbits with low angular momentum, marking a transition between the two limits displayed in Fig.~\ref{fig:LIF_comparison}.

In general the difference in distribution function will be completely erased by the time the SMBH has grown to a size such that all orbits with periapsis $r_p\leq R_\star $ have been captured by the BH. For a Schwarzschild BH the regrowth necessary for total capture of all such orbits is $\mathcal{L}_\text{max}(R_\star )=R_\star \sqrt{\frac{1}{1-2M/R_\star }-1}=4\,M_f$ where $M_f$ is the final mass after regrowth. In the limit $R_\star \gg2M$ this gives 
\begin{equation}
    M_f={R_\star }\sqrt{\frac{2M}{16(R_\star -2M)}}\approx  M\sqrt{\frac{R_\star }{8M}}\approx 5.5\,M\;.
\end{equation}
It follows then that for a regrowth larger then a factor of 5.5 the spike becomes indistinguishable from one formed through adiabatic growth.

\section{Conclusion}
\label{sec:Conclusion}
We have developed a general method to evolve the relativistic distribution function of collisionless particles across transitions between spherically symmetric, static metrics. This method allows us to study transitions that are non-adiabatic on orbital timescales and complements existing treatments of the adiabatic regime in the literature. Our approach reproduces earlier results, including those obtained in the Newtonian limit (see Appendix~\ref{sec:ICN}), and can be straightforwardly applied to any scenario in which particle geodesics can be deterministically evolved. With additional numerical effort, the method can also be extended to transitions between more general stationary spacetimes.

We have applied this method to study, for the first time within a fully self-consistent general-relativistic framework, how the formation of a SMBH via the direct collapse of a SMS affects the surrounding DM distribution. 
In line with earlier studies, we found a sharp enhancement of the DM density around the BH. The enhancement is milder than in scenarios that assume adiabatic BH formation, because the collapse leads to an enhanced capture of particles with low binding energy and thus to a partial depletion of the spike. At the same time, compared to previous analyses based on the simplifying instantaneous collapse approximation, our more realistic collapse model yields a less suppressed central density. We were also able to analyze the effect not only at the level of the DM density profiles, but directly through the distribution function itself, demonstrating that the post-collapse phase-space structure exhibits a clear depletion of circular and nearly circular orbits near the BH. 

If the BH obtained via the direct collapse undergoes a further phase of growth via adiabatic accretion, the features discussed above may gradually be erased. In particular, we find that for any adiabatic DM spike configuration that has been depleted up to some given radius $r_d$, regardless of the mechanism that has caused the depletion, the modifications of the distribution function are completely erased if the BH growth factor $M_f/M$ is larger than $\sqrt{r_d/8M}$. For smaller growth factors, the distribution function interpolates smoothly between the depleted and fully adiabatic spike cases.

These results on the DM profile and the distribution function have direct implications for future gravitational wave observations, since accretion and dynamical friction induced by DM impact compact object inspirals into SMBHs and modify the associated gravitational-wave dephasing. In upcoming work, we will quantify these signatures and investigate how future EMRI detections with space-based interferometers may be used to probe both the nature of DM and the formation histories of SMBHs.

\textit{Acknowledgements.} 
RC would like to thank Annalisa Celotti for her guidance and encouragement during the development of this work. PU acknowledges support from Research grant {\sl Addressing systematic uncertainties in searches for dark matter}, Grant No. 2022F2843L funded by the Italian Ministry of Education, University and Research (MIUR).
BJK acknowledges support from the \textit{Consolidaci\'on Investigadora} Project \textsc{DarkSpikesGW}, reference CNS2023-144071, financed by MCIN/AEI/10.13039/501100011033 and by the European Union ``NextGenerationEU"/PRTR. BJK also acknowledges support from the project SA101P24 (Junta de Castilla y León).
DG acknowledges support from the  Research grant TAsP (Theoretical Astroparticle Physics) funded by INFN. We gratefully acknowledge the support of the Dutch Research Council (NWO) through an Open Competition Domain Science-M grant, project number OCENW.M.21.375.

\setlength{\bibsep}{4pt}
\bibliography{References}

\begin{thebibliography}{71}%
\makeatletter
\providecommand \@ifxundefined [1]{%
 \@ifx{#1\undefined}
}%
\providecommand \@ifnum [1]{%
 \ifnum #1\expandafter \@firstoftwo
 \else \expandafter \@secondoftwo
 \fi
}%
\providecommand \@ifx [1]{%
 \ifx #1\expandafter \@firstoftwo
 \else \expandafter \@secondoftwo
 \fi
}%
\providecommand \natexlab [1]{#1}%
\providecommand \enquote  [1]{``#1''}%
\providecommand \bibnamefont  [1]{#1}%
\providecommand \bibfnamefont [1]{#1}%
\providecommand \citenamefont [1]{#1}%
\providecommand \href@noop [0]{\@secondoftwo}%
\providecommand \href [0]{\begingroup \@sanitize@url \@href}%
\providecommand \@href[1]{\@@startlink{#1}\@@href}%
\providecommand \@@href[1]{\endgroup#1\@@endlink}%
\providecommand \@sanitize@url [0]{\catcode `\\12\catcode `\$12\catcode `\&12\catcode `\#12\catcode `\^12\catcode `\_12\catcode `\%12\relax}%
\providecommand \@@startlink[1]{}%
\providecommand \@@endlink[0]{}%
\providecommand \url  [0]{\begingroup\@sanitize@url \@url }%
\providecommand \@url [1]{\endgroup\@href {#1}{\urlprefix }}%
\providecommand \urlprefix  [0]{URL }%
\providecommand \Eprint [0]{\href }%
\providecommand \doibase [0]{https://doi.org/}%
\providecommand \selectlanguage [0]{\@gobble}%
\providecommand \bibinfo  [0]{\@secondoftwo}%
\providecommand \bibfield  [0]{\@secondoftwo}%
\providecommand \translation [1]{[#1]}%
\providecommand \BibitemOpen [0]{}%
\providecommand \bibitemStop [0]{}%
\providecommand \bibitemNoStop [0]{.\EOS\space}%
\providecommand \EOS [0]{\spacefactor3000\relax}%
\providecommand \BibitemShut  [1]{\csname bibitem#1\endcsname}%
\let\auto@bib@innerbib\@empty
\bibitem [{\citenamefont {Bertone}\ \emph {et~al.}(2005)\citenamefont {Bertone}, \citenamefont {Hooper},\ and\ \citenamefont {Silk}}]{Bertone:2004pz}%
  \BibitemOpen
  \bibfield  {author} {\bibinfo {author} {\bibfnamefont {G.}~\bibnamefont {Bertone}}, \bibinfo {author} {\bibfnamefont {D.}~\bibnamefont {Hooper}},\ and\ \bibinfo {author} {\bibfnamefont {J.}~\bibnamefont {Silk}},\ }\bibfield  {title} {\bibinfo {title} {{Particle dark matter: Evidence, candidates and constraints}},\ }\href {https://doi.org/10.1016/j.physrep.2004.08.031} {\bibfield  {journal} {\bibinfo  {journal} {Phys. Rept.}\ }\textbf {\bibinfo {volume} {405}},\ \bibinfo {pages} {279} (\bibinfo {year} {2005})},\ \Eprint {https://arxiv.org/abs/hep-ph/0404175} {arXiv:hep-ph/0404175} \BibitemShut {NoStop}%
\bibitem [{\citenamefont {Silk}\ \emph {et~al.}(2010)\citenamefont {Silk} \emph {et~al.}}]{Bertone:2010zza}%
  \BibitemOpen
  \bibfield  {author} {\bibinfo {author} {\bibfnamefont {J.}~\bibnamefont {Silk}} \emph {et~al.},\ }\href {https://doi.org/10.1017/CBO9780511770739} {\emph {\bibinfo {title} {{Particle Dark Matter: Observations, Models and Searches}}}},\ edited by\ \bibinfo {editor} {\bibfnamefont {G.}~\bibnamefont {Bertone}}\ (\bibinfo  {publisher} {Cambridge Univ. Press},\ \bibinfo {address} {Cambridge},\ \bibinfo {year} {2010})\BibitemShut {NoStop}%
\bibitem [{\citenamefont {Bertone}\ and\ \citenamefont {Hooper}(2018)}]{Bertone:2016nfn}%
  \BibitemOpen
  \bibfield  {author} {\bibinfo {author} {\bibfnamefont {G.}~\bibnamefont {Bertone}}\ and\ \bibinfo {author} {\bibfnamefont {D.}~\bibnamefont {Hooper}},\ }\bibfield  {title} {\bibinfo {title} {{History of dark matter}},\ }\href {https://doi.org/10.1103/RevModPhys.90.045002} {\bibfield  {journal} {\bibinfo  {journal} {Rev. Mod. Phys.}\ }\textbf {\bibinfo {volume} {90}},\ \bibinfo {pages} {045002} (\bibinfo {year} {2018})},\ \Eprint {https://arxiv.org/abs/1605.04909} {arXiv:1605.04909 [astro-ph.CO]} \BibitemShut {NoStop}%
\bibitem [{\citenamefont {Arcadi}\ \emph {et~al.}(2018)\citenamefont {Arcadi}, \citenamefont {Dutra}, \citenamefont {Ghosh}, \citenamefont {Lindner}, \citenamefont {Mambrini}, \citenamefont {Pierre}, \citenamefont {Profumo},\ and\ \citenamefont {Queiroz}}]{Arcadi:2017kky}%
  \BibitemOpen
  \bibfield  {author} {\bibinfo {author} {\bibfnamefont {G.}~\bibnamefont {Arcadi}}, \bibinfo {author} {\bibfnamefont {M.}~\bibnamefont {Dutra}}, \bibinfo {author} {\bibfnamefont {P.}~\bibnamefont {Ghosh}}, \bibinfo {author} {\bibfnamefont {M.}~\bibnamefont {Lindner}}, \bibinfo {author} {\bibfnamefont {Y.}~\bibnamefont {Mambrini}}, \bibinfo {author} {\bibfnamefont {M.}~\bibnamefont {Pierre}}, \bibinfo {author} {\bibfnamefont {S.}~\bibnamefont {Profumo}},\ and\ \bibinfo {author} {\bibfnamefont {F.~S.}\ \bibnamefont {Queiroz}},\ }\bibfield  {title} {\bibinfo {title} {{The waning of the WIMP? A review of models, searches, and constraints}},\ }\href {https://doi.org/10.1140/epjc/s10052-018-5662-y} {\bibfield  {journal} {\bibinfo  {journal} {Eur. Phys. J. C}\ }\textbf {\bibinfo {volume} {78}},\ \bibinfo {pages} {203} (\bibinfo {year} {2018})},\ \Eprint {https://arxiv.org/abs/1703.07364} {arXiv:1703.07364 [hep-ph]} \BibitemShut {NoStop}%
\bibitem [{\citenamefont {Buckley}\ and\ \citenamefont {Peter}(2018)}]{Buckley:2017ijx}%
  \BibitemOpen
  \bibfield  {author} {\bibinfo {author} {\bibfnamefont {M.~R.}\ \bibnamefont {Buckley}}\ and\ \bibinfo {author} {\bibfnamefont {A.~H.~G.}\ \bibnamefont {Peter}},\ }\bibfield  {title} {\bibinfo {title} {{Gravitational probes of dark matter physics}},\ }\href {https://doi.org/10.1016/j.physrep.2018.07.003} {\bibfield  {journal} {\bibinfo  {journal} {Phys. Rept.}\ }\textbf {\bibinfo {volume} {761}},\ \bibinfo {pages} {1} (\bibinfo {year} {2018})},\ \Eprint {https://arxiv.org/abs/1712.06615} {arXiv:1712.06615 [astro-ph.CO]} \BibitemShut {NoStop}%
\bibitem [{\citenamefont {Bertone}\ and\ \citenamefont {Tait}(2018)}]{Bertone:2018krk}%
  \BibitemOpen
  \bibfield  {author} {\bibinfo {author} {\bibfnamefont {G.}~\bibnamefont {Bertone}}\ and\ \bibinfo {author} {\bibfnamefont {T.}~\bibnamefont {Tait}, \bibfnamefont {M.~P.}},\ }\bibfield  {title} {\bibinfo {title} {{A new era in the search for dark matter}},\ }\href {https://doi.org/10.1038/s41586-018-0542-z} {\bibfield  {journal} {\bibinfo  {journal} {Nature}\ }\textbf {\bibinfo {volume} {562}},\ \bibinfo {pages} {51} (\bibinfo {year} {2018})},\ \Eprint {https://arxiv.org/abs/1810.01668} {arXiv:1810.01668 [astro-ph.CO]} \BibitemShut {NoStop}%
\bibitem [{\citenamefont {Cirelli}\ \emph {et~al.}(2024)\citenamefont {Cirelli}, \citenamefont {Strumia},\ and\ \citenamefont {Zupan}}]{Cirelli:2024ssz}%
  \BibitemOpen
  \bibfield  {author} {\bibinfo {author} {\bibfnamefont {M.}~\bibnamefont {Cirelli}}, \bibinfo {author} {\bibfnamefont {A.}~\bibnamefont {Strumia}},\ and\ \bibinfo {author} {\bibfnamefont {J.}~\bibnamefont {Zupan}},\ }\href {https://arxiv.org/abs/2406.01705} {\bibinfo {title} {{Dark Matter}}} (\bibinfo {year} {2024}),\ \Eprint {https://arxiv.org/abs/2406.01705} {arXiv:2406.01705 [hep-ph]} \BibitemShut {NoStop}%
\bibitem [{\citenamefont {Dubinski}\ and\ \citenamefont {Carlberg}(1991)}]{Dubinski:1991bm}%
  \BibitemOpen
  \bibfield  {author} {\bibinfo {author} {\bibfnamefont {J.}~\bibnamefont {Dubinski}}\ and\ \bibinfo {author} {\bibfnamefont {R.~G.}\ \bibnamefont {Carlberg}},\ }\bibfield  {title} {\bibinfo {title} {{The Structure of cold dark matter halos}},\ }\href {https://doi.org/10.1086/170451} {\bibfield  {journal} {\bibinfo  {journal} {Astrophys. J.}\ }\textbf {\bibinfo {volume} {378}},\ \bibinfo {pages} {496} (\bibinfo {year} {1991})}\BibitemShut {NoStop}%
\bibitem [{\citenamefont {Navarro}\ \emph {et~al.}(1996)\citenamefont {Navarro}, \citenamefont {Frenk},\ and\ \citenamefont {White}}]{Navarro_1996}%
  \BibitemOpen
  \bibfield  {author} {\bibinfo {author} {\bibfnamefont {J.~F.}\ \bibnamefont {Navarro}}, \bibinfo {author} {\bibfnamefont {C.~S.}\ \bibnamefont {Frenk}},\ and\ \bibinfo {author} {\bibfnamefont {S.~D.~M.}\ \bibnamefont {White}},\ }\bibfield  {title} {\bibinfo {title} {{The Structure of cold dark matter halos}},\ }\href {https://doi.org/10.1086/177173} {\bibfield  {journal} {\bibinfo  {journal} {Astrophys. J.}\ }\textbf {\bibinfo {volume} {462}},\ \bibinfo {pages} {563} (\bibinfo {year} {1996})},\ \Eprint {https://arxiv.org/abs/astro-ph/9508025} {arXiv:astro-ph/9508025} \BibitemShut {NoStop}%
\bibitem [{\citenamefont {Straight}\ \emph {et~al.}(2025)\citenamefont {Straight}, \citenamefont {Boylan-Kolchin}, \citenamefont {Bullock}, \citenamefont {Hopkins}, \citenamefont {Shen}, \citenamefont {Necib}, \citenamefont {Lazar}, \citenamefont {Graus},\ and\ \citenamefont {Samuel}}]{Straight:2025udg}%
  \BibitemOpen
  \bibfield  {author} {\bibinfo {author} {\bibfnamefont {M.~C.}\ \bibnamefont {Straight}}, \bibinfo {author} {\bibfnamefont {M.}~\bibnamefont {Boylan-Kolchin}}, \bibinfo {author} {\bibfnamefont {J.~S.}\ \bibnamefont {Bullock}}, \bibinfo {author} {\bibfnamefont {P.~F.}\ \bibnamefont {Hopkins}}, \bibinfo {author} {\bibfnamefont {X.}~\bibnamefont {Shen}}, \bibinfo {author} {\bibfnamefont {L.}~\bibnamefont {Necib}}, \bibinfo {author} {\bibfnamefont {A.}~\bibnamefont {Lazar}}, \bibinfo {author} {\bibfnamefont {A.~S.}\ \bibnamefont {Graus}},\ and\ \bibinfo {author} {\bibfnamefont {J.}~\bibnamefont {Samuel}},\ }\bibfield  {title} {\bibinfo {title} {{Central densities of dark matter halos in FIRE-2 simulations of low-mass galaxies with cold dark matter and self-interacting dark matter}},\ }\href {https://doi.org/10.1093/mnras/staf1539} {\bibfield  {journal} {\bibinfo  {journal} {Mon. Not. Roy. Astron. Soc.}\ }\textbf {\bibinfo {volume} {543}},\ \bibinfo {pages} {1995} (\bibinfo {year} {2025})},\ \Eprint
  {https://arxiv.org/abs/2501.16602} {arXiv:2501.16602 [astro-ph.GA]} \BibitemShut {NoStop}%
\bibitem [{\citenamefont {Ludlow}\ \emph {et~al.}(2013)\citenamefont {Ludlow}, \citenamefont {Navarro}, \citenamefont {Boylan-Kolchin}, \citenamefont {Bett}, \citenamefont {Angulo}, \citenamefont {Li}, \citenamefont {White}, \citenamefont {Frenk},\ and\ \citenamefont {Springel}}]{Ludlow:2013bd}%
  \BibitemOpen
  \bibfield  {author} {\bibinfo {author} {\bibfnamefont {A.~D.}\ \bibnamefont {Ludlow}}, \bibinfo {author} {\bibfnamefont {J.~F.}\ \bibnamefont {Navarro}}, \bibinfo {author} {\bibfnamefont {M.}~\bibnamefont {Boylan-Kolchin}}, \bibinfo {author} {\bibfnamefont {P.~E.}\ \bibnamefont {Bett}}, \bibinfo {author} {\bibfnamefont {R.~E.}\ \bibnamefont {Angulo}}, \bibinfo {author} {\bibfnamefont {M.}~\bibnamefont {Li}}, \bibinfo {author} {\bibfnamefont {S.~D.~M.}\ \bibnamefont {White}}, \bibinfo {author} {\bibfnamefont {C.}~\bibnamefont {Frenk}},\ and\ \bibinfo {author} {\bibfnamefont {V.}~\bibnamefont {Springel}},\ }\bibfield  {title} {\bibinfo {title} {{The Mass Profile and Accretion History of Cold Dark Matter Halos}},\ }\href {https://doi.org/10.1093/mnras/stt526} {\bibfield  {journal} {\bibinfo  {journal} {Mon. Not. Roy. Astron. Soc.}\ }\textbf {\bibinfo {volume} {432}},\ \bibinfo {pages} {1103} (\bibinfo {year} {2013})},\ \Eprint {https://arxiv.org/abs/1302.0288} {arXiv:1302.0288 [astro-ph.CO]} \BibitemShut
  {NoStop}%
\bibitem [{\citenamefont {Schaller}\ \emph {et~al.}(2015)\citenamefont {Schaller}, \citenamefont {Frenk}, \citenamefont {Bower}, \citenamefont {Theuns}, \citenamefont {Jenkins}, \citenamefont {Schaye}, \citenamefont {Crain}, \citenamefont {Furlong}, \citenamefont {Vecchia},\ and\ \citenamefont {McCarthy}}]{Schaller:2014uwa}%
  \BibitemOpen
  \bibfield  {author} {\bibinfo {author} {\bibfnamefont {M.}~\bibnamefont {Schaller}}, \bibinfo {author} {\bibfnamefont {C.~S.}\ \bibnamefont {Frenk}}, \bibinfo {author} {\bibfnamefont {R.~G.}\ \bibnamefont {Bower}}, \bibinfo {author} {\bibfnamefont {T.}~\bibnamefont {Theuns}}, \bibinfo {author} {\bibfnamefont {A.}~\bibnamefont {Jenkins}}, \bibinfo {author} {\bibfnamefont {J.}~\bibnamefont {Schaye}}, \bibinfo {author} {\bibfnamefont {R.~A.}\ \bibnamefont {Crain}}, \bibinfo {author} {\bibfnamefont {M.}~\bibnamefont {Furlong}}, \bibinfo {author} {\bibfnamefont {C.~D.}\ \bibnamefont {Vecchia}},\ and\ \bibinfo {author} {\bibfnamefont {I.~G.}\ \bibnamefont {McCarthy}},\ }\bibfield  {title} {\bibinfo {title} {{Baryon effects on the internal structure of {\ensuremath{\Lambda}}CDM haloes in the EAGLE simulations}},\ }\href {https://doi.org/10.1093/mnras/stv1067} {\bibfield  {journal} {\bibinfo  {journal} {Mon. Not. Roy. Astron. Soc.}\ }\textbf {\bibinfo {volume} {451}},\ \bibinfo {pages} {1247} (\bibinfo {year}
  {2015})},\ \Eprint {https://arxiv.org/abs/1409.8617} {arXiv:1409.8617 [astro-ph.CO]} \BibitemShut {NoStop}%
\bibitem [{\citenamefont {Correa}\ \emph {et~al.}(2015)\citenamefont {Correa}, \citenamefont {Wyithe}, \citenamefont {Schaye},\ and\ \citenamefont {Duffy}}]{Correa_2015}%
  \BibitemOpen
  \bibfield  {author} {\bibinfo {author} {\bibfnamefont {C.~A.}\ \bibnamefont {Correa}}, \bibinfo {author} {\bibfnamefont {J.~S.~B.}\ \bibnamefont {Wyithe}}, \bibinfo {author} {\bibfnamefont {J.}~\bibnamefont {Schaye}},\ and\ \bibinfo {author} {\bibfnamefont {A.~R.}\ \bibnamefont {Duffy}},\ }\bibfield  {title} {\bibinfo {title} {{The accretion history of dark matter haloes {\textendash} III. A physical model for the concentration{\textendash}mass relation}},\ }\href {https://doi.org/10.1093/mnras/stv1363} {\bibfield  {journal} {\bibinfo  {journal} {Mon. Not. Roy. Astron. Soc.}\ }\textbf {\bibinfo {volume} {452}},\ \bibinfo {pages} {1217} (\bibinfo {year} {2015})},\ \Eprint {https://arxiv.org/abs/1502.00391} {arXiv:1502.00391 [astro-ph.CO]} \BibitemShut {NoStop}%
\bibitem [{\citenamefont {Calore}\ \emph {et~al.}(2015)\citenamefont {Calore}, \citenamefont {Bozorgnia}, \citenamefont {Lovell}, \citenamefont {Bertone}, \citenamefont {Schaller}, \citenamefont {Frenk}, \citenamefont {Crain}, \citenamefont {Schaye}, \citenamefont {Theuns},\ and\ \citenamefont {Trayford}}]{Calore:2015oya}%
  \BibitemOpen
  \bibfield  {author} {\bibinfo {author} {\bibfnamefont {F.}~\bibnamefont {Calore}}, \bibinfo {author} {\bibfnamefont {N.}~\bibnamefont {Bozorgnia}}, \bibinfo {author} {\bibfnamefont {M.}~\bibnamefont {Lovell}}, \bibinfo {author} {\bibfnamefont {G.}~\bibnamefont {Bertone}}, \bibinfo {author} {\bibfnamefont {M.}~\bibnamefont {Schaller}}, \bibinfo {author} {\bibfnamefont {C.~S.}\ \bibnamefont {Frenk}}, \bibinfo {author} {\bibfnamefont {R.~A.}\ \bibnamefont {Crain}}, \bibinfo {author} {\bibfnamefont {J.}~\bibnamefont {Schaye}}, \bibinfo {author} {\bibfnamefont {T.}~\bibnamefont {Theuns}},\ and\ \bibinfo {author} {\bibfnamefont {J.~W.}\ \bibnamefont {Trayford}},\ }\bibfield  {title} {\bibinfo {title} {{Simulated Milky Way analogues: implications for dark matter indirect searches}},\ }\href {https://doi.org/10.1088/1475-7516/2015/12/053} {\bibfield  {journal} {\bibinfo  {journal} {JCAP}\ }\textbf {\bibinfo {volume} {12}},\ \bibinfo {pages} {053}},\ \Eprint {https://arxiv.org/abs/1509.02164} {arXiv:1509.02164
  [astro-ph.GA]} \BibitemShut {NoStop}%
\bibitem [{\citenamefont {Schaller}\ \emph {et~al.}(2016)\citenamefont {Schaller} \emph {et~al.}}]{Schaller:2015mua}%
  \BibitemOpen
  \bibfield  {author} {\bibinfo {author} {\bibfnamefont {M.}~\bibnamefont {Schaller}} \emph {et~al.},\ }\bibfield  {title} {\bibinfo {title} {{Dark matter annihilation radiation in hydrodynamic simulations of Milky Way haloes}},\ }\href {https://doi.org/10.1093/mnras/stv2667} {\bibfield  {journal} {\bibinfo  {journal} {Mon. Not. Roy. Astron. Soc.}\ }\textbf {\bibinfo {volume} {455}},\ \bibinfo {pages} {4442} (\bibinfo {year} {2016})},\ \Eprint {https://arxiv.org/abs/1509.02166} {arXiv:1509.02166 [astro-ph.CO]} \BibitemShut {NoStop}%
\bibitem [{\citenamefont {Gondolo}\ and\ \citenamefont {Silk}(1999)}]{Gondolo_1999}%
  \BibitemOpen
  \bibfield  {author} {\bibinfo {author} {\bibfnamefont {P.}~\bibnamefont {Gondolo}}\ and\ \bibinfo {author} {\bibfnamefont {J.}~\bibnamefont {Silk}},\ }\bibfield  {title} {\bibinfo {title} {{Dark matter annihilation at the galactic center}},\ }\href {https://doi.org/10.1103/PhysRevLett.83.1719} {\bibfield  {journal} {\bibinfo  {journal} {Phys. Rev. Lett.}\ }\textbf {\bibinfo {volume} {83}},\ \bibinfo {pages} {1719} (\bibinfo {year} {1999})},\ \Eprint {https://arxiv.org/abs/astro-ph/9906391} {arXiv:astro-ph/9906391} \BibitemShut {NoStop}%
\bibitem [{\citenamefont {Ullio}\ \emph {et~al.}(2001)\citenamefont {Ullio}, \citenamefont {Zhao},\ and\ \citenamefont {Kamionkowski}}]{Ullio_2001}%
  \BibitemOpen
  \bibfield  {author} {\bibinfo {author} {\bibfnamefont {P.}~\bibnamefont {Ullio}}, \bibinfo {author} {\bibfnamefont {H.}~\bibnamefont {Zhao}},\ and\ \bibinfo {author} {\bibfnamefont {M.}~\bibnamefont {Kamionkowski}},\ }\bibfield  {title} {\bibinfo {title} {{A Dark matter spike at the galactic center?}},\ }\href {https://doi.org/10.1103/PhysRevD.64.043504} {\bibfield  {journal} {\bibinfo  {journal} {Phys. Rev. D}\ }\textbf {\bibinfo {volume} {64}},\ \bibinfo {pages} {043504} (\bibinfo {year} {2001})},\ \Eprint {https://arxiv.org/abs/astro-ph/0101481} {arXiv:astro-ph/0101481} \BibitemShut {NoStop}%
\bibitem [{\citenamefont {Bertone}(2024)}]{BERTONE2024116487}%
  \BibitemOpen
  \bibfield  {author} {\bibinfo {author} {\bibfnamefont {G.}~\bibnamefont {Bertone}},\ }\bibfield  {title} {\bibinfo {title} {{Dark matter, black holes, and gravitational waves}},\ }\href {https://doi.org/10.1016/j.nuclphysb.2024.116487} {\bibfield  {journal} {\bibinfo  {journal} {Nucl. Phys. B}\ }\textbf {\bibinfo {volume} {1003}},\ \bibinfo {pages} {116487} (\bibinfo {year} {2024})},\ \Eprint {https://arxiv.org/abs/2404.11513} {arXiv:2404.11513 [astro-ph.CO]} \BibitemShut {NoStop}%
\bibitem [{\citenamefont {Barack}\ \emph {et~al.}(2019)\citenamefont {Barack} \emph {et~al.}}]{Barack:2018yly}%
  \BibitemOpen
  \bibfield  {author} {\bibinfo {author} {\bibfnamefont {L.}~\bibnamefont {Barack}} \emph {et~al.},\ }\bibfield  {title} {\bibinfo {title} {{Black holes, gravitational waves and fundamental physics: a roadmap}},\ }\href {https://doi.org/10.1088/1361-6382/ab0587} {\bibfield  {journal} {\bibinfo  {journal} {Class. Quant. Grav.}\ }\textbf {\bibinfo {volume} {36}},\ \bibinfo {pages} {143001} (\bibinfo {year} {2019})},\ \Eprint {https://arxiv.org/abs/1806.05195} {arXiv:1806.05195 [gr-qc]} \BibitemShut {NoStop}%
\bibitem [{\citenamefont {Berry}\ \emph {et~al.}(2019)\citenamefont {Berry}, \citenamefont {Hughes}, \citenamefont {Sopuerta}, \citenamefont {Chua}, \citenamefont {Heffernan}, \citenamefont {Holley-Bockelmann}, \citenamefont {Mihaylov}, \citenamefont {Miller},\ and\ \citenamefont {Sesana}}]{Berry:2019wgg}%
  \BibitemOpen
  \bibfield  {author} {\bibinfo {author} {\bibfnamefont {C.~P.~L.}\ \bibnamefont {Berry}}, \bibinfo {author} {\bibfnamefont {S.~A.}\ \bibnamefont {Hughes}}, \bibinfo {author} {\bibfnamefont {C.~F.}\ \bibnamefont {Sopuerta}}, \bibinfo {author} {\bibfnamefont {A.~J.~K.}\ \bibnamefont {Chua}}, \bibinfo {author} {\bibfnamefont {A.}~\bibnamefont {Heffernan}}, \bibinfo {author} {\bibfnamefont {K.}~\bibnamefont {Holley-Bockelmann}}, \bibinfo {author} {\bibfnamefont {D.~P.}\ \bibnamefont {Mihaylov}}, \bibinfo {author} {\bibfnamefont {M.~C.}\ \bibnamefont {Miller}},\ and\ \bibinfo {author} {\bibfnamefont {A.}~\bibnamefont {Sesana}},\ }\bibfield  {title} {\bibinfo {title} {{The unique potential of extreme mass-ratio inspirals for gravitational-wave astronomy}},\ }\href@noop {} {\bibfield  {journal} {\bibinfo  {journal} {Bull. Am. Astron. Soc.}\ }\textbf {\bibinfo {volume} {51}},\ \bibinfo {pages} {42} (\bibinfo {year} {2019})},\ \Eprint {https://arxiv.org/abs/1903.03686} {arXiv:1903.03686 [astro-ph.HE]} \BibitemShut
  {NoStop}%
\bibitem [{\citenamefont {Seoane}\ \emph {et~al.}(2023)\citenamefont {Seoane} \emph {et~al.}}]{LISA:2022yao}%
  \BibitemOpen
  \bibfield  {author} {\bibinfo {author} {\bibfnamefont {P.~A.}\ \bibnamefont {Seoane}} \emph {et~al.} (\bibinfo {collaboration} {LISA}),\ }\bibfield  {title} {\bibinfo {title} {{Astrophysics with the Laser Interferometer Space Antenna}},\ }\href {https://doi.org/10.1007/s41114-022-00041-y} {\bibfield  {journal} {\bibinfo  {journal} {Living Rev. Rel.}\ }\textbf {\bibinfo {volume} {26}},\ \bibinfo {pages} {2} (\bibinfo {year} {2023})},\ \Eprint {https://arxiv.org/abs/2203.06016} {arXiv:2203.06016 [gr-qc]} \BibitemShut {NoStop}%
\bibitem [{\citenamefont {Eda}\ \emph {et~al.}(2013)\citenamefont {Eda}, \citenamefont {Itoh}, \citenamefont {Kuroyanagi},\ and\ \citenamefont {Silk}}]{Eda2013}%
  \BibitemOpen
  \bibfield  {author} {\bibinfo {author} {\bibfnamefont {K.}~\bibnamefont {Eda}}, \bibinfo {author} {\bibfnamefont {Y.}~\bibnamefont {Itoh}}, \bibinfo {author} {\bibfnamefont {S.}~\bibnamefont {Kuroyanagi}},\ and\ \bibinfo {author} {\bibfnamefont {J.}~\bibnamefont {Silk}},\ }\bibfield  {title} {\bibinfo {title} {New probe of dark-matter properties: Gravitational waves from an intermediate-mass black hole embedded in a dark-matter minispike},\ }\href {https://doi.org/10.1103/PhysRevLett.110.221101} {\bibfield  {journal} {\bibinfo  {journal} {Physical Review Letters}\ }\textbf {\bibinfo {volume} {110}},\ \bibinfo {pages} {221101} (\bibinfo {year} {2013})},\ \Eprint {https://arxiv.org/abs/1301.5971} {arXiv:1301.5971 [gr-qc]} \BibitemShut {NoStop}%
\bibitem [{\citenamefont {Eda}\ \emph {et~al.}(2015)\citenamefont {Eda}, \citenamefont {Itoh}, \citenamefont {Kuroyanagi},\ and\ \citenamefont {Silk}}]{Eda2015}%
  \BibitemOpen
  \bibfield  {author} {\bibinfo {author} {\bibfnamefont {K.}~\bibnamefont {Eda}}, \bibinfo {author} {\bibfnamefont {Y.}~\bibnamefont {Itoh}}, \bibinfo {author} {\bibfnamefont {S.}~\bibnamefont {Kuroyanagi}},\ and\ \bibinfo {author} {\bibfnamefont {J.}~\bibnamefont {Silk}},\ }\bibfield  {title} {\bibinfo {title} {Gravitational waves as a probe of dark matter minispikes},\ }\href {https://doi.org/10.1103/PhysRevD.91.044045} {\bibfield  {journal} {\bibinfo  {journal} {Physical Review D}\ }\textbf {\bibinfo {volume} {91}},\ \bibinfo {pages} {044045} (\bibinfo {year} {2015})},\ \Eprint {https://arxiv.org/abs/1408.3534} {arXiv:1408.3534 [gr-qc]} \BibitemShut {NoStop}%
\bibitem [{\citenamefont {Macedo}\ \emph {et~al.}(2013)\citenamefont {Macedo}, \citenamefont {Pani}, \citenamefont {Cardoso},\ and\ \citenamefont {Crispino}}]{Macedo2013}%
  \BibitemOpen
  \bibfield  {author} {\bibinfo {author} {\bibfnamefont {C.~F.~B.}\ \bibnamefont {Macedo}}, \bibinfo {author} {\bibfnamefont {P.}~\bibnamefont {Pani}}, \bibinfo {author} {\bibfnamefont {V.}~\bibnamefont {Cardoso}},\ and\ \bibinfo {author} {\bibfnamefont {L.~C.~B.}\ \bibnamefont {Crispino}},\ }\bibfield  {title} {\bibinfo {title} {Into the lair: gravitational-wave signatures of dark matter},\ }\href {https://doi.org/10.1088/0004-637X/774/1/48} {\bibfield  {journal} {\bibinfo  {journal} {Astrophysical Journal}\ }\textbf {\bibinfo {volume} {774}},\ \bibinfo {pages} {48} (\bibinfo {year} {2013})},\ \Eprint {https://arxiv.org/abs/1302.2646} {arXiv:1302.2646 [gr-qc]} \BibitemShut {NoStop}%
\bibitem [{\citenamefont {Barausse}\ \emph {et~al.}(2014)\citenamefont {Barausse}, \citenamefont {Cardoso},\ and\ \citenamefont {Pani}}]{Barausse2014}%
  \BibitemOpen
  \bibfield  {author} {\bibinfo {author} {\bibfnamefont {E.}~\bibnamefont {Barausse}}, \bibinfo {author} {\bibfnamefont {V.}~\bibnamefont {Cardoso}},\ and\ \bibinfo {author} {\bibfnamefont {P.}~\bibnamefont {Pani}},\ }\bibfield  {title} {\bibinfo {title} {Can environmental effects spoil precision gravitational-wave astrophysics?},\ }\href {https://doi.org/10.1103/PhysRevD.89.104059} {\bibfield  {journal} {\bibinfo  {journal} {Physical Review D}\ }\textbf {\bibinfo {volume} {89}},\ \bibinfo {pages} {104059} (\bibinfo {year} {2014})},\ \Eprint {https://arxiv.org/abs/1404.7149} {arXiv:1404.7149 [gr-qc]} \BibitemShut {NoStop}%
\bibitem [{\citenamefont {Yue}\ and\ \citenamefont {Cao}(2019)}]{Yue2019}%
  \BibitemOpen
  \bibfield  {author} {\bibinfo {author} {\bibfnamefont {X.-J.}\ \bibnamefont {Yue}}\ and\ \bibinfo {author} {\bibfnamefont {Z.}~\bibnamefont {Cao}},\ }\bibfield  {title} {\bibinfo {title} {Dark matter minispike: A significant enhancement of eccentricity for intermediate-mass-ratio inspirals},\ }\href {https://doi.org/10.1103/PhysRevD.100.043013} {\bibfield  {journal} {\bibinfo  {journal} {Physical Review D}\ }\textbf {\bibinfo {volume} {100}},\ \bibinfo {pages} {043013} (\bibinfo {year} {2019})},\ \Eprint {https://arxiv.org/abs/1908.10241} {arXiv:1908.10241 [astro-ph.HE]} \BibitemShut {NoStop}%
\bibitem [{\citenamefont {Kavanagh}\ \emph {et~al.}(2020)\citenamefont {Kavanagh}, \citenamefont {Nichols}, \citenamefont {Bertone},\ and\ \citenamefont {Gaggero}}]{Kavanagh2020}%
  \BibitemOpen
  \bibfield  {author} {\bibinfo {author} {\bibfnamefont {B.~J.}\ \bibnamefont {Kavanagh}}, \bibinfo {author} {\bibfnamefont {D.~A.}\ \bibnamefont {Nichols}}, \bibinfo {author} {\bibfnamefont {G.}~\bibnamefont {Bertone}},\ and\ \bibinfo {author} {\bibfnamefont {D.}~\bibnamefont {Gaggero}},\ }\bibfield  {title} {\bibinfo {title} {Detecting dark matter around black holes with gravitational waves: Effects of dark-matter dynamics on the gravitational waveform},\ }\href {https://doi.org/10.1103/PhysRevD.102.083006} {\bibfield  {journal} {\bibinfo  {journal} {Physical Review D}\ }\textbf {\bibinfo {volume} {102}},\ \bibinfo {pages} {083006} (\bibinfo {year} {2020})},\ \Eprint {https://arxiv.org/abs/2002.12811} {arXiv:2002.12811 [gr-qc]} \BibitemShut {NoStop}%
\bibitem [{\citenamefont {Becker}\ \emph {et~al.}(2022)\citenamefont {Becker}, \citenamefont {Sagunski}, \citenamefont {Prinz},\ and\ \citenamefont {Rastgoo}}]{Becker2022}%
  \BibitemOpen
  \bibfield  {author} {\bibinfo {author} {\bibfnamefont {N.}~\bibnamefont {Becker}}, \bibinfo {author} {\bibfnamefont {L.}~\bibnamefont {Sagunski}}, \bibinfo {author} {\bibfnamefont {L.}~\bibnamefont {Prinz}},\ and\ \bibinfo {author} {\bibfnamefont {S.}~\bibnamefont {Rastgoo}},\ }\bibfield  {title} {\bibinfo {title} {Circularization versus eccentrification in intermediate mass ratio inspirals inside dark matter spikes},\ }\href {https://doi.org/10.1103/PhysRevD.105.063029} {\bibfield  {journal} {\bibinfo  {journal} {Physical Review D}\ }\textbf {\bibinfo {volume} {105}},\ \bibinfo {pages} {063029} (\bibinfo {year} {2022})},\ \Eprint {https://arxiv.org/abs/2112.09586} {arXiv:2112.09586 [gr-qc]} \BibitemShut {NoStop}%
\bibitem [{\citenamefont {Speeney}\ \emph {et~al.}(2022)\citenamefont {Speeney}, \citenamefont {Antonelli}, \citenamefont {Baibhav},\ and\ \citenamefont {Berti}}]{Speeney2022}%
  \BibitemOpen
  \bibfield  {author} {\bibinfo {author} {\bibfnamefont {N.}~\bibnamefont {Speeney}}, \bibinfo {author} {\bibfnamefont {A.}~\bibnamefont {Antonelli}}, \bibinfo {author} {\bibfnamefont {V.}~\bibnamefont {Baibhav}},\ and\ \bibinfo {author} {\bibfnamefont {E.}~\bibnamefont {Berti}},\ }\bibfield  {title} {\bibinfo {title} {Impact of relativistic corrections on the detectability of dark-matter spikes with gravitational waves},\ }\href {https://doi.org/10.1103/PhysRevD.106.044027} {\bibfield  {journal} {\bibinfo  {journal} {Physical Review D}\ }\textbf {\bibinfo {volume} {106}},\ \bibinfo {pages} {044027} (\bibinfo {year} {2022})},\ \Eprint {https://arxiv.org/abs/2204.12508} {arXiv:2204.12508 [gr-qc]} \BibitemShut {NoStop}%
\bibitem [{\citenamefont {Cardoso}\ \emph {et~al.}(2022)\citenamefont {Cardoso}, \citenamefont {Destounis}, \citenamefont {Duque}, \citenamefont {Panosso~Macedo},\ and\ \citenamefont {Maselli}}]{Cardoso2022}%
  \BibitemOpen
  \bibfield  {author} {\bibinfo {author} {\bibfnamefont {V.}~\bibnamefont {Cardoso}}, \bibinfo {author} {\bibfnamefont {K.}~\bibnamefont {Destounis}}, \bibinfo {author} {\bibfnamefont {F.}~\bibnamefont {Duque}}, \bibinfo {author} {\bibfnamefont {R.}~\bibnamefont {Panosso~Macedo}},\ and\ \bibinfo {author} {\bibfnamefont {A.}~\bibnamefont {Maselli}},\ }\bibfield  {title} {\bibinfo {title} {Gravitational waves from extreme-mass-ratio systems in astrophysical environments},\ }\href {https://doi.org/10.1103/PhysRevLett.129.241103} {\bibfield  {journal} {\bibinfo  {journal} {Physical Review Letters}\ }\textbf {\bibinfo {volume} {129}},\ \bibinfo {pages} {241103} (\bibinfo {year} {2022})},\ \Eprint {https://arxiv.org/abs/2210.01133} {arXiv:2210.01133 [gr-qc]} \BibitemShut {NoStop}%
\bibitem [{\citenamefont {Nichols}\ \emph {et~al.}(2023)\citenamefont {Nichols}, \citenamefont {Wade},\ and\ \citenamefont {Grant}}]{Nichols2023}%
  \BibitemOpen
  \bibfield  {author} {\bibinfo {author} {\bibfnamefont {D.~A.}\ \bibnamefont {Nichols}}, \bibinfo {author} {\bibfnamefont {B.~A.}\ \bibnamefont {Wade}},\ and\ \bibinfo {author} {\bibfnamefont {A.~M.}\ \bibnamefont {Grant}},\ }\bibfield  {title} {\bibinfo {title} {Secondary accretion of dark matter in intermediate mass-ratio inspirals: Dark-matter dynamics and gravitational-wave phase},\ }\href {https://doi.org/10.1103/PhysRevD.108.124062} {\bibfield  {journal} {\bibinfo  {journal} {Physical Review D}\ }\textbf {\bibinfo {volume} {108}},\ \bibinfo {pages} {124062} (\bibinfo {year} {2023})},\ \Eprint {https://arxiv.org/abs/2309.06498} {arXiv:2309.06498 [gr-qc]} \BibitemShut {NoStop}%
\bibitem [{\citenamefont {Karydas}\ \emph {et~al.}(2025{\natexlab{a}})\citenamefont {Karydas}, \citenamefont {Kavanagh},\ and\ \citenamefont {Bertone}}]{Karydas_2025a}%
  \BibitemOpen
  \bibfield  {author} {\bibinfo {author} {\bibfnamefont {T.~K.}\ \bibnamefont {Karydas}}, \bibinfo {author} {\bibfnamefont {B.~J.}\ \bibnamefont {Kavanagh}},\ and\ \bibinfo {author} {\bibfnamefont {G.}~\bibnamefont {Bertone}},\ }\bibfield  {title} {\bibinfo {title} {{Sharpening the dark matter signature in gravitational waveforms.~I.~Accretion and eccentricity evolution}},\ }\href {https://doi.org/10.1103/PhysRevD.111.063070} {\bibfield  {journal} {\bibinfo  {journal} {Phys. Rev. D}\ }\textbf {\bibinfo {volume} {111}},\ \bibinfo {pages} {063070} (\bibinfo {year} {2025}{\natexlab{a}})},\ \Eprint {https://arxiv.org/abs/2402.13053} {arXiv:2402.13053 [gr-qc]} \BibitemShut {NoStop}%
\bibitem [{\citenamefont {Kavanagh}\ \emph {et~al.}(2025)\citenamefont {Kavanagh}, \citenamefont {Karydas}, \citenamefont {Bertone}, \citenamefont {Di~Cintio},\ and\ \citenamefont {Pasquato}}]{Kavanagh_2025}%
  \BibitemOpen
  \bibfield  {author} {\bibinfo {author} {\bibfnamefont {B.~J.}\ \bibnamefont {Kavanagh}}, \bibinfo {author} {\bibfnamefont {T.~K.}\ \bibnamefont {Karydas}}, \bibinfo {author} {\bibfnamefont {G.}~\bibnamefont {Bertone}}, \bibinfo {author} {\bibfnamefont {P.}~\bibnamefont {Di~Cintio}},\ and\ \bibinfo {author} {\bibfnamefont {M.}~\bibnamefont {Pasquato}},\ }\bibfield  {title} {\bibinfo {title} {{Sharpening the dark matter signature in gravitational waveforms. II. Numerical simulations}},\ }\href {https://doi.org/10.1103/PhysRevD.111.063071} {\bibfield  {journal} {\bibinfo  {journal} {Phys. Rev. D}\ }\textbf {\bibinfo {volume} {111}},\ \bibinfo {pages} {063071} (\bibinfo {year} {2025})},\ \Eprint {https://arxiv.org/abs/2402.13762} {arXiv:2402.13762 [gr-qc]} \BibitemShut {NoStop}%
\bibitem [{\citenamefont {Cardoso}\ and\ \citenamefont {Maselli}(2020)}]{Cardoso2020}%
  \BibitemOpen
  \bibfield  {author} {\bibinfo {author} {\bibfnamefont {V.}~\bibnamefont {Cardoso}}\ and\ \bibinfo {author} {\bibfnamefont {A.}~\bibnamefont {Maselli}},\ }\bibfield  {title} {\bibinfo {title} {Constraints on the astrophysical environment of binaries with gravitational-wave observations},\ }\href {https://doi.org/10.1051/0004-6361/202038010} {\bibfield  {journal} {\bibinfo  {journal} {Astronomy \& Astrophysics}\ }\textbf {\bibinfo {volume} {644}},\ \bibinfo {pages} {A147} (\bibinfo {year} {2020})},\ \Eprint {https://arxiv.org/abs/1909.05870} {arXiv:1909.05870 [astro-ph.HE]} \BibitemShut {NoStop}%
\bibitem [{\citenamefont {Coogan}\ \emph {et~al.}(2022)\citenamefont {Coogan}, \citenamefont {Bertone}, \citenamefont {Gaggero}, \citenamefont {Kavanagh},\ and\ \citenamefont {Nichols}}]{Coogan2022}%
  \BibitemOpen
  \bibfield  {author} {\bibinfo {author} {\bibfnamefont {A.}~\bibnamefont {Coogan}}, \bibinfo {author} {\bibfnamefont {G.}~\bibnamefont {Bertone}}, \bibinfo {author} {\bibfnamefont {D.}~\bibnamefont {Gaggero}}, \bibinfo {author} {\bibfnamefont {B.~J.}\ \bibnamefont {Kavanagh}},\ and\ \bibinfo {author} {\bibfnamefont {D.~A.}\ \bibnamefont {Nichols}},\ }\bibfield  {title} {\bibinfo {title} {Measuring the dark matter environments of black hole binaries with gravitational waves},\ }\href {https://doi.org/10.1103/PhysRevD.105.043009} {\bibfield  {journal} {\bibinfo  {journal} {Physical Review D}\ }\textbf {\bibinfo {volume} {105}},\ \bibinfo {pages} {043009} (\bibinfo {year} {2022})},\ \Eprint {https://arxiv.org/abs/2108.04154} {arXiv:2108.04154 [gr-qc]} \BibitemShut {NoStop}%
\bibitem [{\citenamefont {Cole}\ \emph {et~al.}(2023{\natexlab{a}})\citenamefont {Cole}, \citenamefont {Coogan}, \citenamefont {Kavanagh},\ and\ \citenamefont {Bertone}}]{Cole2023}%
  \BibitemOpen
  \bibfield  {author} {\bibinfo {author} {\bibfnamefont {P.~S.}\ \bibnamefont {Cole}}, \bibinfo {author} {\bibfnamefont {A.}~\bibnamefont {Coogan}}, \bibinfo {author} {\bibfnamefont {B.~J.}\ \bibnamefont {Kavanagh}},\ and\ \bibinfo {author} {\bibfnamefont {G.}~\bibnamefont {Bertone}},\ }\bibfield  {title} {\bibinfo {title} {Measuring dark matter spikes around primordial black holes with einstein telescope and cosmic explorer},\ }\href {https://doi.org/10.1103/PhysRevD.107.083006} {\bibfield  {journal} {\bibinfo  {journal} {Physical Review D}\ }\textbf {\bibinfo {volume} {107}},\ \bibinfo {pages} {083006} (\bibinfo {year} {2023}{\natexlab{a}})},\ \Eprint {https://arxiv.org/abs/2207.07576} {arXiv:2207.07576 [astro-ph.CO]} \BibitemShut {NoStop}%
\bibitem [{\citenamefont {Cole}\ \emph {et~al.}(2023{\natexlab{b}})\citenamefont {Cole}, \citenamefont {Bertone}, \citenamefont {Coogan}, \citenamefont {Gaggero}, \citenamefont {Karydas}, \citenamefont {Kavanagh}, \citenamefont {Spieksma},\ and\ \citenamefont {Tomaselli}}]{Cole2023Nature}%
  \BibitemOpen
  \bibfield  {author} {\bibinfo {author} {\bibfnamefont {P.~S.}\ \bibnamefont {Cole}}, \bibinfo {author} {\bibfnamefont {G.}~\bibnamefont {Bertone}}, \bibinfo {author} {\bibfnamefont {A.}~\bibnamefont {Coogan}}, \bibinfo {author} {\bibfnamefont {D.}~\bibnamefont {Gaggero}}, \bibinfo {author} {\bibfnamefont {T.}~\bibnamefont {Karydas}}, \bibinfo {author} {\bibfnamefont {B.~J.}\ \bibnamefont {Kavanagh}}, \bibinfo {author} {\bibfnamefont {T.~F.~M.}\ \bibnamefont {Spieksma}},\ and\ \bibinfo {author} {\bibfnamefont {G.~M.}\ \bibnamefont {Tomaselli}},\ }\bibfield  {title} {\bibinfo {title} {Distinguishing environmental effects on binary black hole gravitational waveforms},\ }\href {https://doi.org/10.1038/s41550-023-01949-y} {\bibfield  {journal} {\bibinfo  {journal} {Nature Astronomy}\ }\textbf {\bibinfo {volume} {7}},\ \bibinfo {pages} {943} (\bibinfo {year} {2023}{\natexlab{b}})},\ \Eprint {https://arxiv.org/abs/2211.01362} {arXiv:2211.01362 [gr-qc]} \BibitemShut {NoStop}%
\bibitem [{\citenamefont {Hannuksela}\ \emph {et~al.}(2020)\citenamefont {Hannuksela}, \citenamefont {Ng},\ and\ \citenamefont {Li}}]{Hannuksela2020}%
  \BibitemOpen
  \bibfield  {author} {\bibinfo {author} {\bibfnamefont {O.~A.}\ \bibnamefont {Hannuksela}}, \bibinfo {author} {\bibfnamefont {K.~C.~Y.}\ \bibnamefont {Ng}},\ and\ \bibinfo {author} {\bibfnamefont {T.~G.~F.}\ \bibnamefont {Li}},\ }\bibfield  {title} {\bibinfo {title} {Extreme dark matter tests with extreme mass ratio inspirals},\ }\href {https://doi.org/10.1103/PhysRevD.102.103022} {\bibfield  {journal} {\bibinfo  {journal} {Physical Review D}\ }\textbf {\bibinfo {volume} {102}},\ \bibinfo {pages} {103022} (\bibinfo {year} {2020})},\ \Eprint {https://arxiv.org/abs/1906.11845} {arXiv:1906.11845 [astro-ph.CO]} \BibitemShut {NoStop}%
\bibitem [{\citenamefont {Vicente}\ \emph {et~al.}(2025)\citenamefont {Vicente}, \citenamefont {Karydas},\ and\ \citenamefont {Bertone}}]{Vicente:2025gsg}%
  \BibitemOpen
  \bibfield  {author} {\bibinfo {author} {\bibfnamefont {R.}~\bibnamefont {Vicente}}, \bibinfo {author} {\bibfnamefont {T.~K.}\ \bibnamefont {Karydas}},\ and\ \bibinfo {author} {\bibfnamefont {G.}~\bibnamefont {Bertone}},\ }\href@noop {} {\bibinfo {title} {{A fully relativistic treatment of EMRIs in collisionless environments}}} (\bibinfo {year} {2025}),\ \Eprint {https://arxiv.org/abs/2505.09715} {arXiv:2505.09715 [gr-qc]} \BibitemShut {NoStop}%
\bibitem [{\citenamefont {Karydas}\ \emph {et~al.}(2025{\natexlab{b}})\citenamefont {Karydas}, \citenamefont {Vicente},\ and\ \citenamefont {Bertone}}]{Karydas:2025bkj}%
  \BibitemOpen
  \bibfield  {author} {\bibinfo {author} {\bibfnamefont {T.~K.}\ \bibnamefont {Karydas}}, \bibinfo {author} {\bibfnamefont {R.}~\bibnamefont {Vicente}},\ and\ \bibinfo {author} {\bibfnamefont {G.}~\bibnamefont {Bertone}},\ }\href@noop {} {\bibinfo {title} {{Mass and spin coevolution of black holes inspiralling through dark matter}}} (\bibinfo {year} {2025}{\natexlab{b}}),\ \Eprint {https://arxiv.org/abs/2510.13604} {arXiv:2510.13604 [gr-qc]} \BibitemShut {NoStop}%
\bibitem [{\citenamefont {Sadeghian}\ \emph {et~al.}(2013)\citenamefont {Sadeghian}, \citenamefont {Ferrer},\ and\ \citenamefont {Will}}]{Sadeghian_2013}%
  \BibitemOpen
  \bibfield  {author} {\bibinfo {author} {\bibfnamefont {L.}~\bibnamefont {Sadeghian}}, \bibinfo {author} {\bibfnamefont {F.}~\bibnamefont {Ferrer}},\ and\ \bibinfo {author} {\bibfnamefont {C.~M.}\ \bibnamefont {Will}},\ }\bibfield  {title} {\bibinfo {title} {{Dark matter distributions around massive black holes: A general relativistic analysis}},\ }\href {https://doi.org/10.1103/PhysRevD.88.063522} {\bibfield  {journal} {\bibinfo  {journal} {Phys. Rev. D}\ }\textbf {\bibinfo {volume} {88}},\ \bibinfo {pages} {063522} (\bibinfo {year} {2013})},\ \Eprint {https://arxiv.org/abs/1305.2619} {arXiv:1305.2619 [astro-ph.GA]} \BibitemShut {NoStop}%
\bibitem [{\citenamefont {Ferrer}\ \emph {et~al.}(2017)\citenamefont {Ferrer}, \citenamefont {da~Rosa},\ and\ \citenamefont {Will}}]{Ferrer:2017xwm}%
  \BibitemOpen
  \bibfield  {author} {\bibinfo {author} {\bibfnamefont {F.}~\bibnamefont {Ferrer}}, \bibinfo {author} {\bibfnamefont {A.~M.}\ \bibnamefont {da~Rosa}},\ and\ \bibinfo {author} {\bibfnamefont {C.~M.}\ \bibnamefont {Will}},\ }\bibfield  {title} {\bibinfo {title} {{Dark matter spikes in the vicinity of Kerr black holes}},\ }\href {https://doi.org/10.1103/PhysRevD.96.083014} {\bibfield  {journal} {\bibinfo  {journal} {Phys. Rev. D}\ }\textbf {\bibinfo {volume} {96}},\ \bibinfo {pages} {083014} (\bibinfo {year} {2017})},\ \Eprint {https://arxiv.org/abs/1707.06302} {arXiv:1707.06302 [astro-ph.CO]} \BibitemShut {NoStop}%
\bibitem [{\citenamefont {Volonteri}\ \emph {et~al.}(2021)\citenamefont {Volonteri}, \citenamefont {Habouzit},\ and\ \citenamefont {Colpi}}]{Volonteri:2021sfo}%
  \BibitemOpen
  \bibfield  {author} {\bibinfo {author} {\bibfnamefont {M.}~\bibnamefont {Volonteri}}, \bibinfo {author} {\bibfnamefont {M.}~\bibnamefont {Habouzit}},\ and\ \bibinfo {author} {\bibfnamefont {M.}~\bibnamefont {Colpi}},\ }\bibfield  {title} {\bibinfo {title} {{The origins of massive black holes}},\ }\href {https://doi.org/10.1038/s42254-021-00364-9} {\bibfield  {journal} {\bibinfo  {journal} {Nature Rev. Phys.}\ }\textbf {\bibinfo {volume} {3}},\ \bibinfo {pages} {732} (\bibinfo {year} {2021})},\ \Eprint {https://arxiv.org/abs/2110.10175} {arXiv:2110.10175 [astro-ph.GA]} \BibitemShut {NoStop}%
\bibitem [{\citenamefont {{Goulding}}\ and\ \citenamefont {all}(2023)}]{2023ApJ...955L..24G}%
  \BibitemOpen
  \bibfield  {author} {\bibinfo {author} {\bibfnamefont {A.~D.}\ \bibnamefont {{Goulding}}}\ and\ \bibinfo {author} {\bibnamefont {all}},\ }\bibfield  {title} {\bibinfo {title} {{UNCOVER: The Growth of the First Massive Black Holes from JWST/NIRSpec-Spectroscopic Redshift Confirmation of an X-Ray Luminous AGN at z = 10.1}},\ }\href {https://doi.org/10.3847/2041-8213/acf7c5} {\bibfield  {journal} {\bibinfo  {journal} {Astrophysical Journal Letters}\ }\textbf {\bibinfo {volume} {955}},\ \bibinfo {eid} {L24} (\bibinfo {year} {2023})},\ \Eprint {https://arxiv.org/abs/2308.02750} {arXiv:2308.02750 [astro-ph.GA]} \BibitemShut {NoStop}%
\bibitem [{\citenamefont {Latif}\ \emph {et~al.}(2014)\citenamefont {Latif}, \citenamefont {Bovino}, \citenamefont {Van~Borm}, \citenamefont {Grassi}, \citenamefont {Schleicher},\ and\ \citenamefont {Spaans}}]{Latif:2014cda}%
  \BibitemOpen
  \bibfield  {author} {\bibinfo {author} {\bibfnamefont {M.~A.}\ \bibnamefont {Latif}}, \bibinfo {author} {\bibfnamefont {S.}~\bibnamefont {Bovino}}, \bibinfo {author} {\bibfnamefont {C.}~\bibnamefont {Van~Borm}}, \bibinfo {author} {\bibfnamefont {T.}~\bibnamefont {Grassi}}, \bibinfo {author} {\bibfnamefont {D.~R.~G.}\ \bibnamefont {Schleicher}},\ and\ \bibinfo {author} {\bibfnamefont {M.}~\bibnamefont {Spaans}},\ }\bibfield  {title} {\bibinfo {title} {{A UV flux constraint on the formation of direct collapse black holes}},\ }\href {https://doi.org/10.1093/mnras/stu1230} {\bibfield  {journal} {\bibinfo  {journal} {Mon. Not. Roy. Astron. Soc.}\ }\textbf {\bibinfo {volume} {443}},\ \bibinfo {pages} {1979} (\bibinfo {year} {2014})},\ \Eprint {https://arxiv.org/abs/1404.5773} {arXiv:1404.5773 [astro-ph.GA]} \BibitemShut {NoStop}%
\bibitem [{\citenamefont {Chon}\ \emph {et~al.}(2018{\natexlab{a}})\citenamefont {Chon}, \citenamefont {Hosokawa},\ and\ \citenamefont {Yoshida}}]{Chon:2018SMS}%
  \BibitemOpen
  \bibfield  {author} {\bibinfo {author} {\bibfnamefont {S.}~\bibnamefont {Chon}}, \bibinfo {author} {\bibfnamefont {T.}~\bibnamefont {Hosokawa}},\ and\ \bibinfo {author} {\bibfnamefont {N.}~\bibnamefont {Yoshida}},\ }\bibfield  {title} {\bibinfo {title} {Radiation hydrodynamics simulations of the formation of direct-collapse supermassive stellar systems},\ }\href {https://doi.org/10.1093/mnras/sty086} {\bibfield  {journal} {\bibinfo  {journal} {Monthly Notices of the Royal Astronomical Society}\ }\textbf {\bibinfo {volume} {475}},\ \bibinfo {pages} {4104–4121} (\bibinfo {year} {2018}{\natexlab{a}})}\BibitemShut {NoStop}%
\bibitem [{\citenamefont {Bertone}\ \emph {et~al.}(2025)\citenamefont {Bertone}, \citenamefont {Wierda}, \citenamefont {Gaggero}, \citenamefont {Kavanagh}, \citenamefont {Volonteri},\ and\ \citenamefont {Yoshida}}]{Bertone_2025}%
  \BibitemOpen
  \bibfield  {author} {\bibinfo {author} {\bibfnamefont {G.}~\bibnamefont {Bertone}}, \bibinfo {author} {\bibfnamefont {A.~R. A.~C.}\ \bibnamefont {Wierda}}, \bibinfo {author} {\bibfnamefont {D.}~\bibnamefont {Gaggero}}, \bibinfo {author} {\bibfnamefont {B.~J.}\ \bibnamefont {Kavanagh}}, \bibinfo {author} {\bibfnamefont {M.}~\bibnamefont {Volonteri}},\ and\ \bibinfo {author} {\bibfnamefont {N.}~\bibnamefont {Yoshida}},\ }\bibfield  {title} {\bibinfo {title} {{Toward a realistic description of dark matter overdensities around black holes}},\ }\href {https://doi.org/10.1103/5nnf-8fz9} {\bibfield  {journal} {\bibinfo  {journal} {Phys. Rev. D}\ }\textbf {\bibinfo {volume} {112}},\ \bibinfo {pages} {043537} (\bibinfo {year} {2025})},\ \Eprint {https://arxiv.org/abs/2404.08731} {arXiv:2404.08731 [astro-ph.CO]} \BibitemShut {NoStop}%
\bibitem [{\citenamefont {Lindquist}(1966)}]{Lindquist:1966igj}%
  \BibitemOpen
  \bibfield  {author} {\bibinfo {author} {\bibfnamefont {R.~W.}\ \bibnamefont {Lindquist}},\ }\bibfield  {title} {\bibinfo {title} {{Relativistic transport theory}},\ }\href {https://doi.org/10.1016/0003-4916(66)90207-7} {\bibfield  {journal} {\bibinfo  {journal} {Annals Phys.}\ }\textbf {\bibinfo {volume} {37}},\ \bibinfo {pages} {487} (\bibinfo {year} {1966})}\BibitemShut {NoStop}%
\bibitem [{\citenamefont {{Rasio}}\ \emph {et~al.}(1989)\citenamefont {{Rasio}}, \citenamefont {{Shapiro}},\ and\ \citenamefont {{Teukolsky}}}]{RST}%
  \BibitemOpen
  \bibfield  {author} {\bibinfo {author} {\bibfnamefont {F.~A.}\ \bibnamefont {{Rasio}}}, \bibinfo {author} {\bibfnamefont {S.~L.}\ \bibnamefont {{Shapiro}}},\ and\ \bibinfo {author} {\bibfnamefont {S.~A.}\ \bibnamefont {{Teukolsky}}},\ }\bibfield  {title} {\bibinfo {title} {{Solving the Vlasov Equation in General Relativity}},\ }\href {https://doi.org/10.1086/167785} {\bibfield  {journal} {\bibinfo  {journal} {\apj}\ }\textbf {\bibinfo {volume} {344}},\ \bibinfo {pages} {146} (\bibinfo {year} {1989})}\BibitemShut {NoStop}%
\bibitem [{\citenamefont {Jeans}(1915)}]{Jeans:1915}%
  \BibitemOpen
  \bibfield  {author} {\bibinfo {author} {\bibfnamefont {J.~H.}\ \bibnamefont {Jeans}},\ }\bibfield  {title} {\bibinfo {title} {{On the theory of star-streaming and the structure of the universe}},\ }\href {https://doi.org/10.1093/mnras/76.2.70} {\bibfield  {journal} {\bibinfo  {journal} {Mon. Not. Roy. Astron. Soc.}\ }\textbf {\bibinfo {volume} {76}},\ \bibinfo {pages} {70} (\bibinfo {year} {1915})}\BibitemShut {NoStop}%
\bibitem [{\citenamefont {{Fackerell}}(1968)}]{Fackerell}%
  \BibitemOpen
  \bibfield  {author} {\bibinfo {author} {\bibfnamefont {E.~D.}\ \bibnamefont {{Fackerell}}},\ }\bibfield  {title} {\bibinfo {title} {{Relativistic Stellar Dynamics}},\ }\href {https://doi.org/10.1086/149693} {\bibfield  {journal} {\bibinfo  {journal} {\apj}\ }\textbf {\bibinfo {volume} {153}},\ \bibinfo {pages} {643} (\bibinfo {year} {1968})}\BibitemShut {NoStop}%
\bibitem [{\citenamefont {Rioseco}\ and\ \citenamefont {Sarbach}(2017)}]{Rioseco_2017}%
  \BibitemOpen
  \bibfield  {author} {\bibinfo {author} {\bibfnamefont {P.}~\bibnamefont {Rioseco}}\ and\ \bibinfo {author} {\bibfnamefont {O.}~\bibnamefont {Sarbach}},\ }\bibfield  {title} {\bibinfo {title} {{Accretion of a relativistic, collisionless kinetic gas into a Schwarzschild black hole}},\ }\href {https://doi.org/10.1088/1361-6382/aa65fa} {\bibfield  {journal} {\bibinfo  {journal} {Class. Quant. Grav.}\ }\textbf {\bibinfo {volume} {34}},\ \bibinfo {pages} {095007} (\bibinfo {year} {2017})},\ \Eprint {https://arxiv.org/abs/1611.02389} {arXiv:1611.02389 [gr-qc]} \BibitemShut {NoStop}%
\bibitem [{\citenamefont {{Binney}}\ and\ \citenamefont {{Tremaine}}(1987)}]{1987gady.book.....B}%
  \BibitemOpen
  \bibfield  {author} {\bibinfo {author} {\bibfnamefont {J.}~\bibnamefont {{Binney}}}\ and\ \bibinfo {author} {\bibfnamefont {S.}~\bibnamefont {{Tremaine}}},\ }\href@noop {} {\emph {\bibinfo {title} {{Galactic dynamics}}}}\ (\bibinfo {year} {1987})\BibitemShut {NoStop}%
\bibitem [{\citenamefont {Hosokawa}\ \emph {et~al.}(2013)\citenamefont {Hosokawa}, \citenamefont {Yorke}, \citenamefont {Inayoshi}, \citenamefont {Omukai},\ and\ \citenamefont {Yoshida}}]{Hosokawa_2013}%
  \BibitemOpen
  \bibfield  {author} {\bibinfo {author} {\bibfnamefont {T.}~\bibnamefont {Hosokawa}}, \bibinfo {author} {\bibfnamefont {H.~W.}\ \bibnamefont {Yorke}}, \bibinfo {author} {\bibfnamefont {K.}~\bibnamefont {Inayoshi}}, \bibinfo {author} {\bibfnamefont {K.}~\bibnamefont {Omukai}},\ and\ \bibinfo {author} {\bibfnamefont {N.}~\bibnamefont {Yoshida}},\ }\bibfield  {title} {\bibinfo {title} {Formation of primordial supermassive stars by rapid mass accretion},\ }\href {https://doi.org/10.1088/0004-637x/778/2/178} {\bibfield  {journal} {\bibinfo  {journal} {The Astrophysical Journal}\ }\textbf {\bibinfo {volume} {778}},\ \bibinfo {pages} {178} (\bibinfo {year} {2013})}\BibitemShut {NoStop}%
\bibitem [{\citenamefont {Chon}\ \emph {et~al.}(2018{\natexlab{b}})\citenamefont {Chon}, \citenamefont {Hosokawa},\ and\ \citenamefont {Yoshida}}]{Chon_2018}%
  \BibitemOpen
  \bibfield  {author} {\bibinfo {author} {\bibfnamefont {S.}~\bibnamefont {Chon}}, \bibinfo {author} {\bibfnamefont {T.}~\bibnamefont {Hosokawa}},\ and\ \bibinfo {author} {\bibfnamefont {N.}~\bibnamefont {Yoshida}},\ }\bibfield  {title} {\bibinfo {title} {Radiation hydrodynamics simulations of the formation of direct-collapse supermassive stellar systems},\ }\href {https://doi.org/10.1093/mnras/sty086} {\bibfield  {journal} {\bibinfo  {journal} {Monthly Notices of the Royal Astronomical Society}\ }\textbf {\bibinfo {volume} {475}},\ \bibinfo {pages} {4104–4121} (\bibinfo {year} {2018}{\natexlab{b}})}\BibitemShut {NoStop}%
\bibitem [{\citenamefont {Umeda}\ \emph {et~al.}(2016)\citenamefont {Umeda}, \citenamefont {Hosokawa}, \citenamefont {Omukai},\ and\ \citenamefont {Yoshida}}]{Umeda_2016}%
  \BibitemOpen
  \bibfield  {author} {\bibinfo {author} {\bibfnamefont {H.}~\bibnamefont {Umeda}}, \bibinfo {author} {\bibfnamefont {T.}~\bibnamefont {Hosokawa}}, \bibinfo {author} {\bibfnamefont {K.}~\bibnamefont {Omukai}},\ and\ \bibinfo {author} {\bibfnamefont {N.}~\bibnamefont {Yoshida}},\ }\bibfield  {title} {\bibinfo {title} {{The Final Fates of Accreting Supermassive Stars}},\ }\href {https://doi.org/10.3847/2041-8205/830/2/L34} {\bibfield  {journal} {\bibinfo  {journal} {Astrophys. J. Lett.}\ }\textbf {\bibinfo {volume} {830}},\ \bibinfo {pages} {L34} (\bibinfo {year} {2016})},\ \Eprint {https://arxiv.org/abs/1609.04457} {arXiv:1609.04457 [astro-ph.SR]} \BibitemShut {NoStop}%
\bibitem [{\citenamefont {Oppenheimer}\ and\ \citenamefont {Volkoff}(1939)}]{Oppenheimer:1939ne}%
  \BibitemOpen
  \bibfield  {author} {\bibinfo {author} {\bibfnamefont {J.~R.}\ \bibnamefont {Oppenheimer}}\ and\ \bibinfo {author} {\bibfnamefont {G.~M.}\ \bibnamefont {Volkoff}},\ }\bibfield  {title} {\bibinfo {title} {{On massive neutron cores}},\ }\href {https://doi.org/10.1103/PhysRev.55.374} {\bibfield  {journal} {\bibinfo  {journal} {Phys. Rev.}\ }\textbf {\bibinfo {volume} {55}},\ \bibinfo {pages} {374} (\bibinfo {year} {1939})}\BibitemShut {NoStop}%
\bibitem [{\citenamefont {{Peebles}}(1972)}]{1972GReGr...3...63P}%
  \BibitemOpen
  \bibfield  {author} {\bibinfo {author} {\bibfnamefont {P.~J.~E.}\ \bibnamefont {{Peebles}}},\ }\bibfield  {title} {\bibinfo {title} {{Gravitational collapse and related phenomena from an empirical point of view, or, black holes are where you find them.}},\ }\href {https://doi.org/10.1007/BF00755923} {\bibfield  {journal} {\bibinfo  {journal} {General Relativity and Gravitation}\ }\textbf {\bibinfo {volume} {3}},\ \bibinfo {pages} {63} (\bibinfo {year} {1972})}\BibitemShut {NoStop}%
\bibitem [{\citenamefont {{Quinlan}}\ \emph {et~al.}(1995)\citenamefont {{Quinlan}}, \citenamefont {{Hernquist}},\ and\ \citenamefont {{Sigurdsson}}}]{1995ApJ...440..554Q}%
  \BibitemOpen
  \bibfield  {author} {\bibinfo {author} {\bibfnamefont {G.~D.}\ \bibnamefont {{Quinlan}}}, \bibinfo {author} {\bibfnamefont {L.}~\bibnamefont {{Hernquist}}},\ and\ \bibinfo {author} {\bibfnamefont {S.}~\bibnamefont {{Sigurdsson}}},\ }\bibfield  {title} {\bibinfo {title} {{Models of Galaxies with Central Black Holes: Adiabatic Growth in Spherical Galaxies}},\ }\href {https://doi.org/10.1086/175295} {\bibfield  {journal} {\bibinfo  {journal} {\apj}\ }\textbf {\bibinfo {volume} {440}},\ \bibinfo {pages} {554} (\bibinfo {year} {1995})},\ \Eprint {https://arxiv.org/abs/astro-ph/9407005} {arXiv:astro-ph/9407005 [astro-ph]} \BibitemShut {NoStop}%
\bibitem [{\citenamefont {Witzany}(2022)}]{witzany2022actionanglecoordinatesblackholegeodesics}%
  \BibitemOpen
  \bibfield  {author} {\bibinfo {author} {\bibfnamefont {V.}~\bibnamefont {Witzany}},\ }\href@noop {} {\bibinfo {title} {{Action-angle coordinates for black-hole geodesics I: Spherically symmetric and Schwarzschild}}} (\bibinfo {year} {2022}),\ \Eprint {https://arxiv.org/abs/2203.11952} {arXiv:2203.11952 [gr-qc]} \BibitemShut {NoStop}%
\bibitem [{\citenamefont {Landau}\ and\ \citenamefont {Lifshitz}(1976)}]{Landau_adiabatic}%
  \BibitemOpen
  \bibfield  {author} {\bibinfo {author} {\bibfnamefont {L.~D.}\ \bibnamefont {Landau}}\ and\ \bibinfo {author} {\bibfnamefont {E.~M.}\ \bibnamefont {Lifshitz}},\ }\bibfield  {title} {\bibinfo {title} {{THE} {CANONICAL} {EQUATIONS}†},\ }in\ \href {https://doi.org/10.1016/B978-0-08-050347-9.50012-5} {\emph {\bibinfo {booktitle} {Mechanics ({Third} {Edition})}}},\ \bibinfo {editor} {edited by\ \bibinfo {editor} {\bibfnamefont {L.~D.}\ \bibnamefont {Landau}}\ and\ \bibinfo {editor} {\bibfnamefont {E.~M.}\ \bibnamefont {Lifshitz}}}\ (\bibinfo  {publisher} {Butterworth-Heinemann},\ \bibinfo {address} {Oxford},\ \bibinfo {year} {1976})\ pp.\ \bibinfo {pages} {131--167}\BibitemShut {NoStop}%
\bibitem [{\citenamefont {Oppenheimer}\ and\ \citenamefont {Snyder}(1939)}]{OS}%
  \BibitemOpen
  \bibfield  {author} {\bibinfo {author} {\bibfnamefont {J.~R.}\ \bibnamefont {Oppenheimer}}\ and\ \bibinfo {author} {\bibfnamefont {H.}~\bibnamefont {Snyder}},\ }\bibfield  {title} {\bibinfo {title} {On continued gravitational contraction},\ }\href {https://doi.org/10.1103/PhysRev.56.455} {\bibfield  {journal} {\bibinfo  {journal} {Phys. Rev.}\ }\textbf {\bibinfo {volume} {56}},\ \bibinfo {pages} {455} (\bibinfo {year} {1939})}\BibitemShut {NoStop}%
\bibitem [{\citenamefont {{Israel}}(1966)}]{Israel1}%
  \BibitemOpen
  \bibfield  {author} {\bibinfo {author} {\bibfnamefont {W.}~\bibnamefont {{Israel}}},\ }\bibfield  {title} {\bibinfo {title} {{Singular hypersurfaces and thin shells in general relativity}},\ }\href {https://doi.org/10.1007/BF02710419} {\bibfield  {journal} {\bibinfo  {journal} {Nuovo Cimento B Serie}\ }\textbf {\bibinfo {volume} {44}},\ \bibinfo {pages} {1} (\bibinfo {year} {1966})}\BibitemShut {NoStop}%
\bibitem [{\citenamefont {Chu}\ and\ \citenamefont {Tan}(2022)}]{chu2021}%
  \BibitemOpen
  \bibfield  {author} {\bibinfo {author} {\bibfnamefont {C.-S.}\ \bibnamefont {Chu}}\ and\ \bibinfo {author} {\bibfnamefont {H.~S.}\ \bibnamefont {Tan}},\ }\href {https://doi.org/10.3390/universe8050250} {\bibinfo {title} {{Generalized Darmois{\textendash}Israel Junction Conditions}}} (\bibinfo {year} {2022}),\ \Eprint {https://arxiv.org/abs/2103.06314} {arXiv:2103.06314 [hep-th]} \BibitemShut {NoStop}%
\bibitem [{\citenamefont {Liouville}(1838)}]{Liouville1838}%
  \BibitemOpen
  \bibfield  {author} {\bibinfo {author} {\bibfnamefont {J.}~\bibnamefont {Liouville}},\ }\bibfield  {title} {\bibinfo {title} {Note sur la théorie de la variation des constantes arbitraires},\ }\href@noop {} {\bibfield  {journal} {\bibinfo  {journal} {\href{https://www.numdam.org/item/JMPA_1838_1_3__342_0/}{Journal de Mathématiques Pures et Appliquées, Serie 1, Volume 3}}\ ,\ \bibinfo {pages} {342}} (\bibinfo {year} {1838})}\BibitemShut {NoStop}%
\bibitem [{\citenamefont {{Lynden-Bell}}(1967)}]{1967MNRAS.136..101L}%
  \BibitemOpen
  \bibfield  {author} {\bibinfo {author} {\bibfnamefont {D.}~\bibnamefont {{Lynden-Bell}}},\ }\bibfield  {title} {\bibinfo {title} {{Statistical mechanics of violent relaxation in stellar systems}},\ }\href {https://doi.org/10.1093/mnras/136.1.101} {\bibfield  {journal} {\bibinfo  {journal} {Monthly Notices of the Royal Astronomical Society}\ }\textbf {\bibinfo {volume} {136}},\ \bibinfo {pages} {101} (\bibinfo {year} {1967})}\BibitemShut {NoStop}%
\bibitem [{\citenamefont {Chavanis}(2002)}]{chavanis2002statisticalmechanicsviolentrelaxation}%
  \BibitemOpen
  \bibfield  {author} {\bibinfo {author} {\bibfnamefont {P.~H.}\ \bibnamefont {Chavanis}},\ }\href@noop {} {\bibinfo {title} {{Statistical mechanics of violent relaxation in stellar systems}}} (\bibinfo {year} {2002}),\ \Eprint {https://arxiv.org/abs/astro-ph/0212205} {arXiv:astro-ph/0212205} \BibitemShut {NoStop}%
\bibitem [{\citenamefont {Tremaine}\ \emph {et~al.}(1986)\citenamefont {Tremaine}, \citenamefont {Hénon},\ and\ \citenamefont {Lynden-Bell}}]{Tremaine1986HFunctions}%
  \BibitemOpen
  \bibfield  {author} {\bibinfo {author} {\bibfnamefont {S.}~\bibnamefont {Tremaine}}, \bibinfo {author} {\bibfnamefont {M.}~\bibnamefont {Hénon}},\ and\ \bibinfo {author} {\bibfnamefont {D.}~\bibnamefont {Lynden-Bell}},\ }\bibfield  {title} {\bibinfo {title} {H-functions and mixing in violent relaxation},\ }\href {https://doi.org/10.1093/mnras/219.2.285} {\bibfield  {journal} {\bibinfo  {journal} {Monthly Notices of the Royal Astronomical Society}\ }\textbf {\bibinfo {volume} {219}},\ \bibinfo {pages} {285} (\bibinfo {year} {1986})}\BibitemShut {NoStop}%
\bibitem [{\citenamefont {Merrall}\ and\ \citenamefont {Henriksen}(2003)}]{Merrall_2003}%
  \BibitemOpen
  \bibfield  {author} {\bibinfo {author} {\bibfnamefont {T.~E.~C.}\ \bibnamefont {Merrall}}\ and\ \bibinfo {author} {\bibfnamefont {R.~N.}\ \bibnamefont {Henriksen}},\ }\bibfield  {title} {\bibinfo {title} {{Relaxation of a collisionless system and the transition to a new equilibrium velocity distribution}},\ }\href {https://doi.org/10.1086/377249} {\bibfield  {journal} {\bibinfo  {journal} {Astrophys. J.}\ }\textbf {\bibinfo {volume} {595}},\ \bibinfo {pages} {43} (\bibinfo {year} {2003})},\ \Eprint {https://arxiv.org/abs/astro-ph/0306268} {arXiv:astro-ph/0306268} \BibitemShut {NoStop}%
\bibitem [{\citenamefont {Rioseco}\ and\ \citenamefont {Sarbach}(2024)}]{RiosecoSarbach2024}%
  \BibitemOpen
  \bibfield  {author} {\bibinfo {author} {\bibfnamefont {P.}~\bibnamefont {Rioseco}}\ and\ \bibinfo {author} {\bibfnamefont {O.}~\bibnamefont {Sarbach}},\ }\bibfield  {title} {\bibinfo {title} {{Phase Space Mixing of a Vlasov Gas in the Exterior of a Kerr Black Hole}},\ }\href {https://doi.org/10.1007/s00220-024-04956-1} {\bibfield  {journal} {\bibinfo  {journal} {Commun. Math. Phys.}\ }\textbf {\bibinfo {volume} {405}},\ \bibinfo {pages} {105} (\bibinfo {year} {2024})},\ \Eprint {https://arxiv.org/abs/2302.12849} {arXiv:2302.12849 [gr-qc]} \BibitemShut {NoStop}%
\bibitem [{\citenamefont {Rioseco}\ and\ \citenamefont {Sarbach}(2020)}]{RiosecoSarbach2020_PhaseSpaceMixing}%
  \BibitemOpen
  \bibfield  {author} {\bibinfo {author} {\bibfnamefont {P.}~\bibnamefont {Rioseco}}\ and\ \bibinfo {author} {\bibfnamefont {O.}~\bibnamefont {Sarbach}},\ }\bibfield  {title} {\bibinfo {title} {{Phase space mixing in an external gravitational central potential}},\ }\href {https://doi.org/10.1088/1361-6382/ababb3} {\bibfield  {journal} {\bibinfo  {journal} {Class. Quant. Grav.}\ }\textbf {\bibinfo {volume} {37}},\ \bibinfo {pages} {195027} (\bibinfo {year} {2020})},\ \Eprint {https://arxiv.org/abs/2005.05988} {arXiv:2005.05988 [gr-qc]} \BibitemShut {NoStop}%
\end{thebibliography}%

\appendix

\section{Instant Collapse in the Newtonian Limit}
\label{sec:ICN}

\begin{figure}
\includegraphics[width=0.45\textwidth]{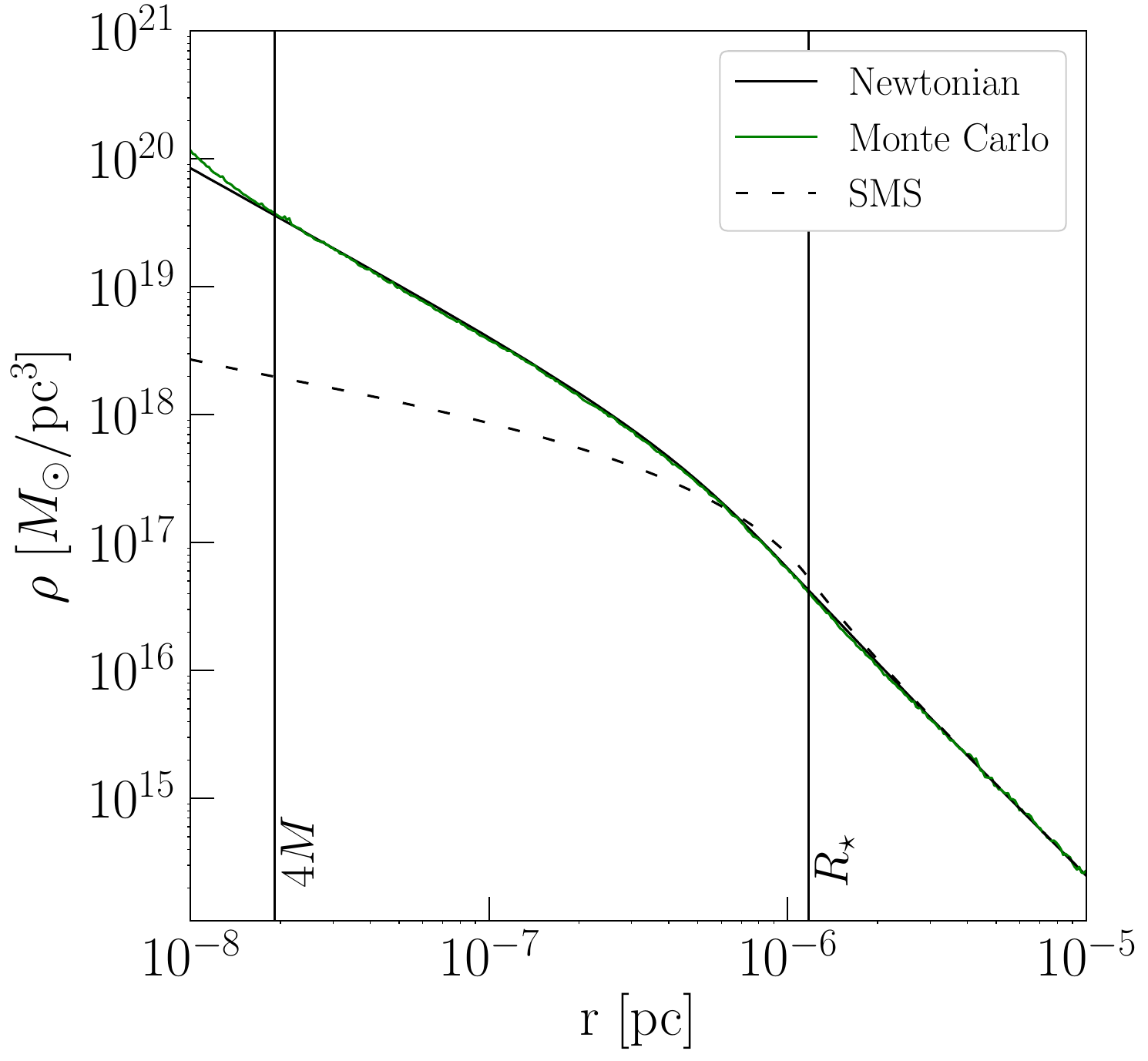}
    \caption{
    Comparison of the DM density computed through the same Monte Carlo procedure as Ref.~\cite{Bertone_2025} and our analytical method both computed in the instant Newtonian case. The dashed line shows the DM density after the adiabatic formation of the SMS, while the solid line shows the density after instantaneous collapse of the star to a BH. The green line shows the Monte Carlo results.}
\label{montecarlo}
\end{figure}

Following the same prescription as in Ref.~\cite{Ullio_2001}, when treating the collapse into a BH as instantaneous in a Newtonian setting, the DM particles conserve their velocity and position at the moment of collapse. The definition of the distribution function in the Newtonian limit is
\begin{equation}
    {f}(\vec{x}, \vec{v})=\frac{\mathrm{d}  N}{\mathrm{d}^3{x}\;\mathrm{d}^3{v}}\,.
\end{equation}
We also know that we can relate the energy before and after collapse by the equation
\begin{equation}
    E_s\left(E_c, r\right)=E_c-\left(\frac{ M}{r}-\Phi_{s}\left(r\right)\right)\,.
\end{equation}  
where we use as in the main paper the subscript $s$ to denote the values before collapse and $c$ for long after collapse. Then using the same reasoning as in the main paper it's straightforward to derive the Newtonian counterpart of Eq.~\eqref{coarsef} as
\begin{equation}
{f}_c(E_c,L)=\oint  \frac{ {f}_s\left(E_s, L\right)}{v_{r}\;T_c} \dd  r\,.
\end{equation}
$T$ here is the Newtonian orbital period of the DM particles. This gives the equation for the post collapse density of DM as:
\begin{equation}
\begin{aligned}
\rho(r)=\frac{4 \pi\mu}{r^2} {\int}\dd E {\int} \mathrm{d}L\;   \frac{L}{\sqrt{2\frac{M}{r}-2E-\frac{L^2}{r^2}}}\\\times \int \mathrm{d}r_c \;\frac{2 {f}_s\left(E_s, L\right)}{T_c\sqrt{2\frac{M}{r_c}-2E-\frac{L^2}{r_c^2}} }\,.
\end{aligned}
\end{equation}
The result computed with this method agrees to high precision with the one obtained through the Monte Carlo procedure (adapted to a fully Newtonian context) from Ref.~\cite{Bertone_2025} as shown in Fig.~\ref{montecarlo}. Note also that if we consider only circular orbits we can express $v_{r}$ and $T_c$ as functions of the initial radius, allowing us to recover the same result as Ref.~\cite{Ullio_2001}:

\begin{equation}
\begin{aligned}
\rho(r)=\frac{4}{r^2}\int \mathrm{d}r_i \frac{r_i^2\rho_i}{T_c(r_i) v_{r}(r_i,r)}\,.
\end{aligned}
\end{equation}

\section{The ``Instantaneous" limit in GR}
\label{sec:ICR}
Of course in GR there is no such thing as an instantaneous collapse as the concept of simultaneity in an extended region of space is frame-dependent. However, in the case of a halo in static equilibrium, there is an interesting choice of frame that recovers the correct Newtonian limit from Ref.~\cite{Ullio_2001}. This is the limit in which the collapse appears as ``instantaneous'' in a stationary reference frame with respect to the DM's particle enesamble which represents the limiting case in which the collapse happens quickly with respect to the movement of particles in the halo and can be considered as the opposite limit to the adiabatic evolution in which the orbital time is much quicker then the timescale over which the metric changes significantly.

In this limit we can use a space-like hypersurface defined by $t=0$ so that on the side of the hypersurface corresponding to $t<0$ the metric is that of the constant density star while for $t>0$ the metric becomes fully Schwarzschild. Note that this type of change in the metric is not mediated by the Einstein equations and is not physical, hence there is no clear path to apply the Israel junction conditions as the induced metrics cannot match. We can however still ask how the geodesics of the DM particles evolve in this limit. Note that for particles outside the star at collapse the metric is unchanged so that the particle will maintain the same properties before and after collapse. The angular momentum of all particles will also stay the same as the discontinuity does not break the spherical symmetry of the system. 

Under this setup the relevant Christoffel symbols for radial and time components will include terms that have delta functions. In our case, this turns into an ODE with distributional coefficients; at $t=0$ there are impulsive terms from the time divergences which do not allow the equation to be solved numerically. Across a small interval $(- \epsilon,\,+\epsilon)$ in affine parameter there is a finite jump in $\dd x^\lambda/\dd\tau$ that depends on the jump in $g_{\alpha\beta}$ in the form
\begin{equation}
\left.\frac{\dd x^\lambda}{\dd\tau}\right|_{\epsilon_+} - \left.\frac{\dd x^\lambda}{\dd\tau}\right|_{\epsilon_-} = - \int_{-\epsilon}^{+\epsilon} \Gamma^\lambda_{\;\alpha\beta}\, \frac{\dd x^\alpha}{\dd\tau}\frac{\dd x^\beta}{\dd\tau}\,\dd\tau\;,
\end{equation} 
In this small interval the distributional limit of the ODE leaves unchanged the angular components of 4-velocity and the coordinate position. It further gives the relation
\begin{equation}
\mathcal{E}_s=\sqrt{\frac{A}{{1-\frac{2M}{r}}}\left({\mathcal{E}_\delta^2}-{(u^r_{\delta})^2}\left(\frac{1-\frac{2m}{r}}{1-\frac{2M}{r}}\right)\right)}\;,
\end{equation}
which can easily be checked to respect the Liouville theorem across the jump in metric at $t=0$. The subscripts $\delta$ are those associated with the ``instant collapse'' considered here. We can therefore again write the final distribution function as
\begin{equation}
{f}_\delta(\mathcal{E_\delta},\mathcal{L})=\frac{1}{{T}_r(\mathcal{E_\delta,\mathcal{L}})}\oint { {f}_s\left(\mathcal{E}_s, \mathcal{L}\right)}\frac{u^t_f}{u_r^f} \;\mathrm{d}r\,.
\end{equation}
\begin{figure}[t!]
\includegraphics[width=0.45\textwidth]{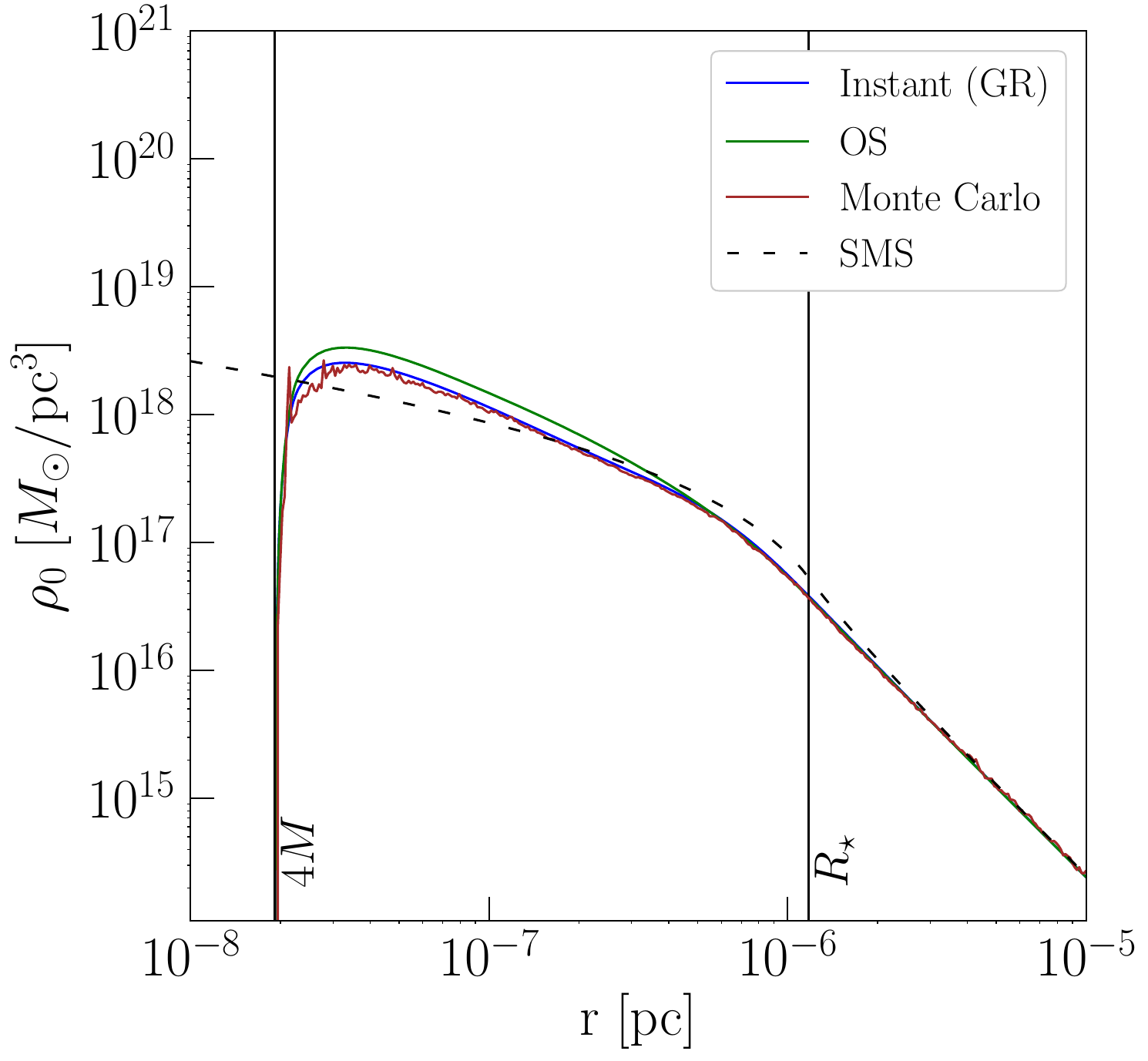}
    \caption{DM rest mass density profiles after the instant collapse and the OS collapse. The result from the same Monte Carlo procedure as described in Ref.~\cite{Bertone_2025} is also shown.}
\label{Newtonian vs GR}
\end{figure}
In Fig.~\ref{Newtonian vs GR}, we show the effects the resulting DM rest mass density in this instantaneous limit. We compare this both to the OS case discussed in the main paper and to the semi-relativistic case from Ref.~\cite{Bertone_2025}. We can see that this limit captures well the results of the semi-relativistic calculation and as expected offers a slightly more extreme suppression of the spike with respect to the OS case. Note further that in the Newtonian Limit this recovers the same result as in Sec.~\ref{sec:ICN}. This is precisely why the result is so close to the one obtained from the methodology used in~\cite{Bertone_2025}; the particle energy is evolved on the surface of simultaneity of the particle ensamble corresponding to the static solution of the distribution function both in the relativistic and Newtonian case. In addition, the majority of particles that survive are those which are far from the BH at collapse, for which the Newtonian limit of the final energy applies. The majority of the GR effects in this case are then restricted to the modification of the DM orbits that pass close to the BH and the subsequent increase in DM captured by the central object.

\section{DM Crossings}
\label{A}
At a given instant $t_{\rm out}$, a DM particle with energy $\mathcal{E}$ and angular momentum~$\mathcal{L}$ crosses the star surface outwards, meaning that $r_{\zeta}(t_{\rm out})=r_\star (t_{\rm out})$ and $\dot{r}_\zeta(t_{\rm out})>\dot{r}_\star(t_{\rm out})$. The radial velocities in coordinate time are
\begin{align}
    \dot{r}_\star&= -\left(1-\frac{2M}{r}\right) \sqrt {1-\frac{1-\frac {2M}{r}}{1-\frac {2M}{R_\star }}}\,, \\
    \dot{r}_\zeta&= \pm \left(1-\frac{2M}{r}\right) \sqrt{1 - \frac{\left(1 - \frac{2M}{r}\right)\left(1 + \frac{\mathcal{L}^{2}}{r^{2}}\right)}{\mathcal{E}^2}}\,,
\end{align}
In the first case, where the particle exits by falling slower then the star's surface ($\dot{r}_\zeta(t_{\rm out})<0$), we have $(1+\mathcal{L}^2/r_{\rm out}^2)>\mathcal{E}^2/(1-2M/R_\star )$. This will also hold for all $r\leq r_{\rm out}$, and thus the coordinate time necessary for the DM particle to reach r will be such that 
\begin{equation}
    \int_{r_{\rm out}}^r \dd r'/\dot{r}_\star<\int_{r_{\rm out}}^r \dd r'/\dot{r}_\zeta\,.
\end{equation} 
With this it is clear that the DM particle will never catch up to the surface.

In the second case, when the DM particle exits the surface with positive radial velocity, it is enough to show that the coordinate time necessary for a DM particle to go from the apoapsis $r_{\rm apo}>r_{\rm out}$ to any $r<r_{\rm out}$ is larger than the time the star surface takes to reach that same $r$. This can easily be seen from
\begin{equation}
\begin{aligned}
    \int_{r_{\rm out}}^r\frac{\dd r}{\dot{r}_\zeta}>\int_r^{r_{\rm apo}} \frac{\sqrt{1 - \frac{2M}{r_{\rm apo}}}\;\dd r'}{\left(1-\frac{2M}{r'}\right) \sqrt{{\frac{2M}{r'} - \frac{2M}{r_{\rm apo}}}}}\,,\\
    \int_r^{r_{\rm out}} \frac{\sqrt{1 - \frac{2M}{r_{\rm out}}}\;\dd r'}{\left(1-\frac{2M}{r'}\right) \sqrt{{\frac{2M}{r'} - \frac{2M}{r_{\rm out}}}}}>\int_{r_{\rm out}}^{r}\frac{\dd r}{\dot{r}_\star}\,,
\end{aligned}
\end{equation}
where we have used the fact that the coordinate time that a radially infalling particle takes to go from $r_{0}$ to $r$, starting with zero velocity, decreases with decreasing $r_{\rm 0}$. 

This is intuitively obvious, but for the interested reader we offer a quick proof below for any $r_0\geq r\geq 3M$, which covers all the orbits relevant to this paper.

What we want to show is that
\begin{equation}
\begin{aligned}
    \frac{\dd}{\dd r_0}\mathcal{T}(r_0,r)=\frac{\dd}{\dd r_0}\int_{r}^{r_0} \frac{\sqrt{1 - \frac{2M}{r_{0}}}\;\dd r'}{\left(1-\frac{2M}{r'}\right) \sqrt{{\frac{2M}{r'} - \frac{2M}{r_{0}}}}}\hspace{16pt} \\=\frac{M}{r_0^2\sqrt{1-\frac{2M}{r_0}}}\int_{r}^{r_0} \frac{\;\dd r'}{\left(1-\frac{2M}{r'}\right) \sqrt{{\frac{2M}{r'} - \frac{2M}{r_{0}}}}}\hspace{28pt} \\+\sqrt{1-\frac{2M}{r_0}}\frac{\dd}{\dd r_0}\int_{r}^{r_0} \frac{\;\dd r'}{\left(1-\frac{2M}{r'}\right) \sqrt{{\frac{2M}{r'} - \frac{2M}{r_{0}}}}}>0 \;.
    \label{this}
\end{aligned}
\end{equation}
The first integral is strictly positive and can be ignored. We can compute the second term by using the substitutions $x=2M /r$, $x'=2M /r'$ and $x_0=2M /r_0$, then integrating by parts to write
\begin{equation}
\begin{aligned}
        \frac{\dd}{\dd r_0}\int_{r}^{r_0} &\frac{\dd r'}{\left(1-\frac{2M}{r'}\right) \sqrt{{\frac{2M}{r'} - \frac{2M}{r_{0}}}}}=
         \\ &\frac{{x_0^2}}{(x^2-x^3)\sqrt{{x - x_0}}}\\ -&x_0^2\;\frac{\dd}{\dd x
        _0}\int_{x}^{x_0} \frac{2(3x'-2)\sqrt{x'-x_0}\;\dd x'}{x'^3\left(1-x'\right)^2 }\,.
        \label{cancel}
\end{aligned}
\end{equation}
The first term is again positive and can be ignored. Then using the Leibniz rule to take the derivative in the second term gives 
\begin{equation}
\begin{aligned}
      -x_0^2\frac{\dd}{\dd x
        _0}\int_{x}^{x_0} \frac{2(3x'-2)\sqrt{x'-x_0}\;\dd x'}{x'^3\left(1-x'\right)^2 }=\\ \int_{x_0}^{x} \frac{-x_0^2(3x'-2)\;\dd x'}{x'^3\left(1-x'\right)^2 \sqrt{x'-x_0}}.
\end{aligned}
\end{equation}   
This integral has a negative contribution only for $x'>2/3$. So for $x\leq2/3$ the derivative is strictly positive and it is manifestly true that, for any $ r\geq 3M$,
\begin{equation}
    \frac{\dd}{\dd r_0}\mathcal{T}(r_0,r)>0\;.
\end{equation}


\end{document}